\documentclass{aa}  
\usepackage{natbib}
\usepackage{graphicx}
\usepackage{txfonts}
\usepackage{hyperref}
\usepackage{url}
\usepackage{color}

\begin{document} 

  \title{ODUSSEAS: Upgraded version with new reference scale and parameter determinations for 82 planet-host M dwarf stars in SWEET-Cat}
  
  \author{A. Antoniadis-Karnavas 
          \inst{1,2}, S. G. Sousa\inst{1}, E. Delgado-Mena\inst{1}, N. C. Santos\inst{1,2}, and D. T. Andreasen\inst{3}}
          
   \institute{Instituto de Astrofísica e Ciências do Espaço, Universidade do Porto, CAUP, Rua das Estrelas, 4150-762 Porto, Portugal \\
  \email{alexandros.antoniadis@astro.up.pt}  
 \and
 Departamento de Física e Astronomia, Faculdade de Ciências, Universidade do Porto, Rua do Campo Alegre, 4169-007 Porto, Portugal 
 \and 
 Independent Researcher
\\}

  \date{Received 14 May 2024 / Accepted 8 July 2024}

  \abstract
  {}
   {Obtaining accurate derivations of stellar atmospheric parameters is crucial in the fields of stellar and exoplanet characterization. We present the upgraded version of our computational tool ODUSSEAS with a new reference scale applied to derive $T_{\mathrm{eff}}$ and [Fe/H] values for M dwarfs.}
   {The new reference dataset of ODUSSEAS consists of $T_{\mathrm{eff}}$ values based on interferometry, and [Fe/H] values derived by applying updated values for the parallaxes. These reference parameters are related to the pseudo-equivalent widths (EWs) of more than 4000 stellar absorption lines. The machine learning Python "scikit learn" package creates models to determine the stellar parameters for subsequent analysis.}
   {We determined $T_{\mathrm{eff}}$ and [Fe/H] values for 82 planet-host stars in SWEET-Cat. We demonstrate that our new version of ODUSSEAS is capable of determining the parameters with a greater accuracy than the original by comparing our results to other methods in literature. We also compared our parameters for the same stars by measuring their spectra obtained from several instruments, showing the consistency of our determinations with standard deviation of 30 K and 0.03 dex. Finally, we examined the correlation among planetary mass and stellar metallicity, confirming prior evidence indicating that massive planets mainly form around metal-rich stars in the case of M dwarfs as well.}
   {}
   {}

\titlerunning{Upgraded ODUSSEAS and parameter determinations for M dwarfs in SWEET-Cat}
\authorrunning{A. Antoniadis-Karnavas et al.}

\maketitle
%________________________________________________________________

\section{Introduction}
\label{intro}

Accurate determinations of stellar parameters are essential for understanding the formation and evolution of stars. In the case of M dwarfs, which are the majority of stars in our Galaxy, their characterization is important to construct the dynamical and chemical evolution of the Galaxy \citep{lepine07, hejazi20}. 
In the fast-growing field of exoplanet detection and characterization, the stars influence the properties of the planets forming and orbiting around them as well as the measurement of their masses and radii \citep{everett13, martinez19, loaiza23}. Abundance determinations of the host stars are important to better understand the formation and evolution of planetary systems \citep{adibekyan13, ghezzi21}.

There are some difficulties that need to be overcome in the derivation of stellar atmospheric parameters for M dwarfs, such as effective temperature ($T_{\mathrm{eff}}$) and metallicity (expressed as [Fe/H]). 
The spectroscopic study of M dwarfs is more complicated compared to that of FGK stars, because their molecules are the dominant sources of opacity \citep{woolf05}, creating thousands of spectral lines that are poorly known and many of them blend with each other. Therefore, the position of the continuum is hardly identified in spectra of M dwarfs. A common approach to derive the atmospheric parameters for FGK stars is by measuring the equivalent widths (EWs) of many metal lines, but in the case of M dwarfs, the measurements of pseudo-EWs have been used \citep{neves12, neves14, maldonado15}.  
Methods using spectral synthesis in the optical have not achieved as precise results as in FGK stars, because of the poor knowledge of many molecular line strengths.  
However, spectral synthesis in the near-infrared (NIR) has made progress, as shown by several studies \citep{one12,lind16,raj18, passegger19}.
Due to the complexity of M dwarf spectra, most attempts to derive $T_{\mathrm{eff}}$ and [Fe/H] have used photometric calibrations \citep{bonfils05,johnsonapps09,neves12} or spectroscopic indices \citep{rojasayala10,rojasayala12,mann13a}. Uncertainties in [Fe/H] vary from 0.20 dex using photometric calibrations to 0.10 dex using spectroscopic scales in the NIR \citep{rojasayala12}. Regarding $T_{\mathrm{eff}}$, precisions better than 100 K have been achieved, but determinations still suffer from uncertainties and systematic errors that range from 150 to 300 K \citep{casagrande08,rojasayala12}.
More accurate methods of determining M dwarf parameters could extend the knowledge and the input for improving the theoretical modeling of stars, which in turn, would lead to an improved understanding and greater progress in stellar characterization. 

Machine learning has become a very popular approach, with applications in many fields of science nowadays. The algorithms of machine learning methods receive input data and by applying statistical analysis they can predict output values accurately, without the need for the user to explicitly create a specific model beforehand.
An overall review of the applications of machine learning to astrophysical problems can be found in \citet{ball10} and \citet{ivezic14}. The interest of creating automatic processes with machine learning algorithms has begun to emerge from the vast volume of survey data \citep{howard17}.
Recently, in the field of stellar astronomy, machine learning methods have been applied to a vast series of problems such as the identification of symbiotic stars \citep{akras19}, the improvement of period-luminosity relations for variable stars \citep{ucci19}, and the determination of mass, age and distance for red-giant stars \citep{das19}. 
In particular, for the characterization of M dwarfs, apart from our original work \citep{antoniadis-karnavas20}, \citet{passegger20} presented a deep neural network to analyze high-resolution stellar spectra and predict stellar atmospheric parameters. The convolutional network was trained on synthetic PHOENIX-ACES spectra in different optical and NIR wavelength regions. 

Given the utility of machine learning techniques and the difficulty in obtaining parameters for M dwarfs, we developed a machine learning method useful for this kind of stars \citep{antoniadis-karnavas20}. Our code, Observing Dwarfs Using Stellar Spectroscopic Energy-Absorption Shapes (ODUSSEAS\footnote{\url{https://github.com/AlexandrosAntoniadis/ODUSSEAS}}), follows the pseudo-EW approach. It is based on a supervised machine learning algorithm, meaning that it is provided with both input and expected output and uses these to create a model. 
It makes use of the machine learning "scikit learn" package of Python and offers a quick automatic derivation of $T_{\mathrm{eff}}$ and [Fe/H] for M dwarf stars using their 1D spectra and resolutions as input. 
The input to the machine learning function of ODUSSEAS are the values of precomputed pseudo-EWs for a set of HARPS GTO spectra and the expected output are the values of their reference $T_{\mathrm{eff}}$ and [Fe/H] from selected literature studies, respectively. 
In the original version of the tool, the reference values of $T_{\mathrm{eff}}$ came from \citet{casagrande08}, following the IRFM method as modified for M dwarfs, and the reference values of [Fe/H] came from \citet{neves12}, for a dataset of 65 stars. 

After the publication of the tool, part of our team participated to a comparative study of methods for determination in the work by \citet{passegger22}. In that study, stellar atmospheric parameters of 18 well-studied M dwarfs observed with the CARMENES spectrograph were determined following different approaches. Four different methods were applied to the spectra:\ two of them, Pass-19 code \citep{passegger18, passegger19} and SteparSyn \citep{tabernero21}, are based on synthetic spectra fitting; while the other two, deep learning \citep{passegger20} and ODUSSEAS \citep{antoniadis-karnavas20}, are based on the machine learning concept. 
In that work, the discrepancies in the derived stellar parameters were analyzed in several analysis runs. The goal was to minimize these discrepancies and find stellar parameters that would be more consistent with the literature value. 
Similarly to the comparison of the reference $T_{\mathrm{eff}}$, with values by \citet{mann15} as showed in \citet{antoniadis-karnavas20}, it was noticed that the photometric reference scale of ODUSSEAS gave results underestimated, compared to the other three methods, as well as to the literature values. 
The mean differences of the ODUSSEAS results were: -86 K to the literature median, -123 K to the interferometric literature values, -178 K to Pass-18 code, -132 K to SteparSyn, and -176 K to deep learning, respectively. 
From these comparisons, it was clear that the IRFM/MOITE-based photometric reference scale was consistently underestimating $T_{\mathrm{eff}}$ values for M dwarfs. Thus, we examined the possibility of replacing our photometric reference dataset with the reference values from interferometry as a potential improvement of our determinations.

\section{Upgrade with new reference scale}
\label{newscale}

To tackle the problem presented in \citep{passegger22}, we searched the literature to find $T_{\mathrm{eff}}$ values based on interferometry for a sufficient number of stars among the 110 we already had HARPS GTO spectra for.

One year after the publication of ODUSSEAS \citep{antoniadis-karnavas20}, \citet {khata21} presented a sample of 23 target stars with high precision long-baseline interferometric measurements of radii \citep{boyajian12, mann15, rabus19} for the calibration of stellar parameters. In that work, 13 calibrators were chosen from \citet{newton15} where the bolometric luminosities were measured from multi-color photometry \citep{boyajian12} and $T_{\mathrm{eff}}$ were calculated using the interferometric radius and bolometric flux \citep{mann13b}. They supplemented this calibration sample with 6 M dwarfs taken from \citet{mann15}, where the bolometric flux was calculated by integrating over the radiative flux density and $T_{\mathrm{eff}}$ was calculated by fitting the optical spectra with the PHOENIX (BT-Settl) models \citep{allard13}. They also added four calibrators from \citet{rabus19}, where the bolometric flux was estimated using PHOENIX models \citep{husser13}, along with photometric observations and $T_{\mathrm{eff}}$ was estimated using the Stefan-Boltzmann law. \citet{khata21} proceeded to a calibration, deriving stellar parameters for 271 stars. 
We had HARPS GTO spectra available for 43 of those 271 stars. 
Moreover, we increased our new reference dataset by adding 4 more stars from the HARPS GTO sample, with values derived by \citet{rabus19}. \citet{rabus19} reported on 13 new high-precision measurements of stellar diameters for low-mass dwarfs obtained by means of NIR long-baseline interferometry with PIONIER at the Very Large Telescope Interferometer (VLTI). Together with accurate parallaxes from Gaia DR2, these measurements provided precise estimates for their linear radii, $T_{\mathrm{eff}}$, masses, and luminosities.
Thus, ODUSSEAS was upgraded by implementing a new reference dataset of 47 stars with $T_{\mathrm{eff}}$ based on interferometry \citep{khata21, rabus19}. 

For these 47 stars, we derived [Fe/H] values using the same method as in the original reference dataset, by applying the following relation from \citet{neves12}: 

\begin{equation}
[Fe/H] = 0.57 \Delta(V - K_s) - 0.17.
\end{equation}

Here, $\Delta(V - K_s)$ = $(V - K_s)_{obs} - (V - K_s)_{iso}$, whereby $(V - K_s)_{obs}$ is the observed $V - K$ color and $(V - K_s)_{iso}$ is a fifth-order polynomial function of the absolute magnitude, $M_K{_s}$, with the coefficients being in increasing order: (51.1413, -39.3756, 12.2862, -1.83916, 0.134266, -0.00382023).
As input values to the equation, we used the $V$ and $K$ magnitudes reported in Simbad\footnote{\url{http://simbad.cds.unistra.fr/simbad/}} and the updated values of parallaxes from Gaia eDR3 \citep{gaia20} for the calculation of $M_K{_s}$.
%Simbad\footnote{\url{http://simbad.cds.unistra.fr/simbad/}}

\begin{figure}
\centering
\includegraphics[width= \hsize]{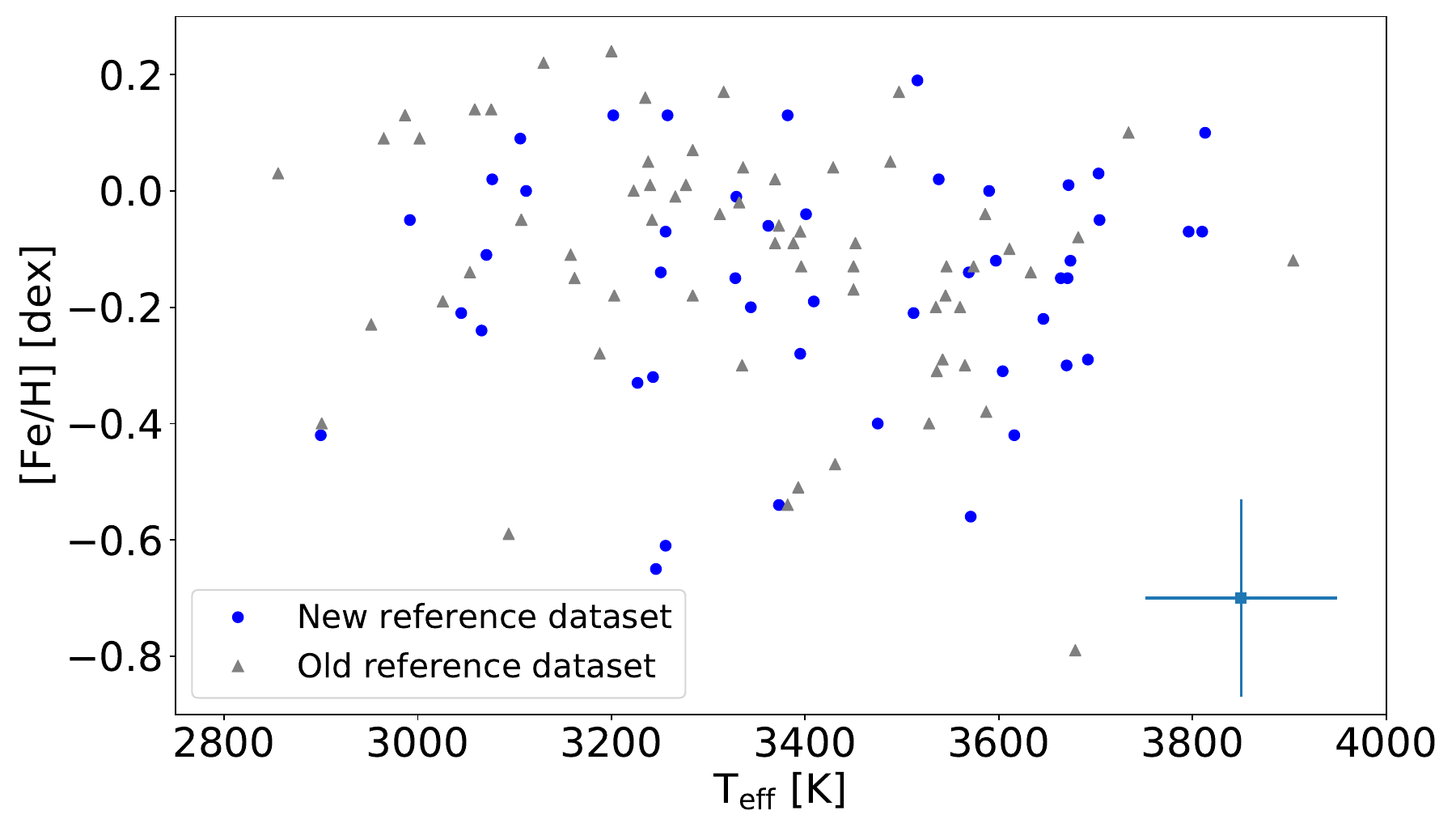} \\
\caption[Distribution of the new reference dataset]{Distribution of the new reference dataset (blue circles) in comparison to the old one (grey triangles). The cross symbol represents the uncertainties of the new reference $T_{\mathrm{eff}}$ and [Fe/H], which are 99 K and 0.17 dex, respectively.}
\label{newrange} 
\end{figure}

\begin{figure}
\centering
$\begin{array}{c}
\includegraphics[width= \hsize]{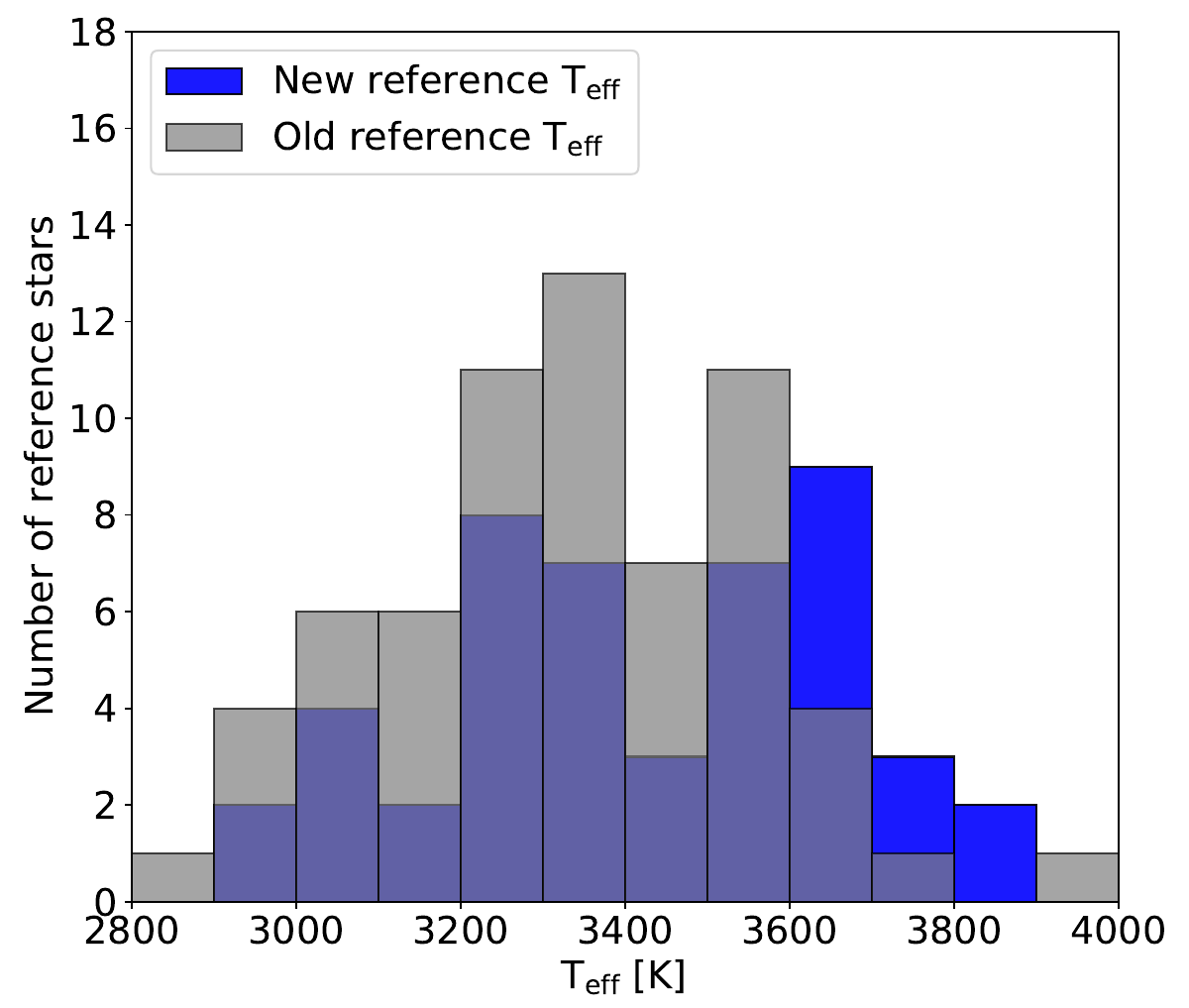} \\
\includegraphics[width= \hsize]{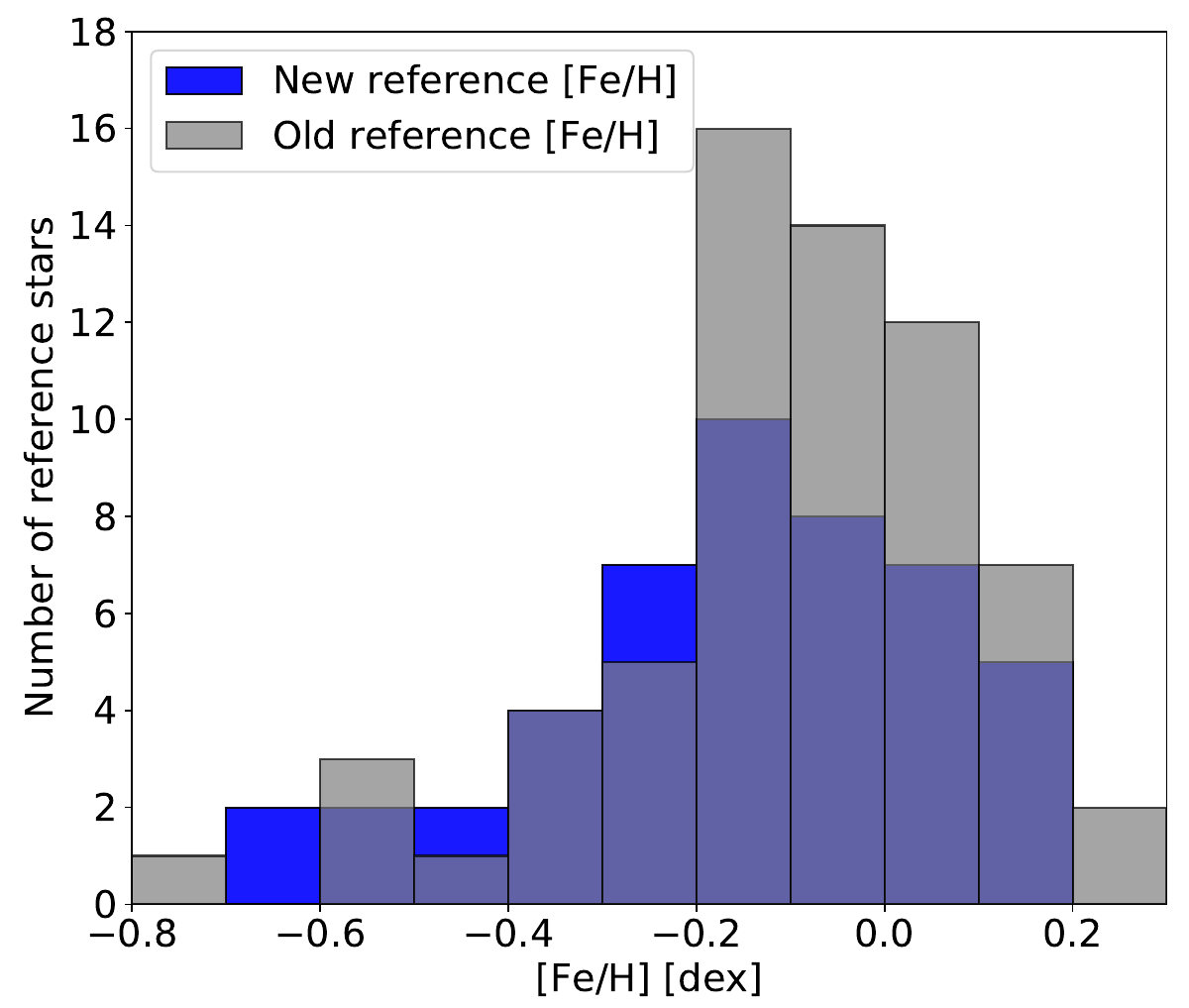}\\
\end{array}$
\caption[Histograms of the new and old reference dataset]{Histograms for $T_{\mathrm{eff}}$ (upper panel) and [Fe/H] (lower panel) of the new (blue) and old (grey) reference datasets.}
\label{new_old_reference_histograms}
\end{figure}

The new reference dataset is presented in Table~\ref{upgraderefparam}.
The methods of derivation for the new reference values have intrinsic uncertainties of 99 K for $T_{\mathrm{eff}}$ and 0.17 dex for [Fe/H], from \citet{khata21} and \citet{neves12}, respectively.
The distribution of the new reference dataset is presented in Fig~\ref{newrange}, compared to the old one. The new reference values range from 2900 to 3813 K for $T_{\mathrm{eff}}$ and from -0.65 to 0.19 dex for [Fe/H].
The histograms of the new reference dataset, compared to the old one, are presented in Fig~\ref{new_old_reference_histograms}.
The parameter space of applicability can be considered to range from 2700 to 4000 K for $T_{\mathrm{eff}}$ and from -0.82 to 0.24 dex for [Fe/H]. 
We have tested the new reference dataset on stars with parameters around these extreme values and we obtained accurate determinations with similar levels of precision to those of determinations that are inside the range of the reference parameters.
The tool can operate beyond the strict range of reference parameters, as its machine learning models extrapolate reliably to the limits mentioned above.

Among the 37 stars in common across both reference datasets, the average difference between the new reference and the old reference scale is 151 K for $T_{\mathrm{eff}}$ and -0.01 dex for [Fe/H].
As we present and discuss below, these average differences between the reference datasets translate to average differences of the same order in the determinations for several samples of stars. 
Users can select (in the options) the scale whose reference dataset will then be applied in determining the parameters.

Similar steps as those applied to the original reference dataset presented by \citet{antoniadis-karnavas20}, were followed again, in order to check the machine learning efficiency, accuracy, and precision with this new reference dataset. 
Since our new reference dataset consists of fewer stars (47) than the original one (65), we decided to increase the percentage of stars in the training group from 70\% to 80\%, in order to include a relatively greater number of stars in the process of building the machine learning models, while still keeping a sufficient number of stars (10) for testing the predictions and measuring the mean absolute errors of the new models.

Again, we calculated  the machine-learning regression metrics of "explained variance" (E.V.) and "r2" scores, as well as the mean absolute errors of the models, for our new reference dataset.
The E.V. score varies mostly between 0.89 and 0.90, while the r2 score is between 0.88 and 0.89. These values are slightly lower than the respective scores between 0.92 and 0.94 of the original reference dataset, because the new population of reference stars is smaller than the initial one (47 compared to 65 stars). Still, the new scores are quite high for making predictions with sufficient accuracy and precision regarding M dwarfs.

The mean absolute errors of the machine learning models are around 65 K for $T_{\mathrm{eff}}$ and 0.04 dex for [Fe/H] for all resolutions. Compared to the model errors of the original reference dataset (around 30 K and 0.04 dex, respectively), the precision of $T_{\mathrm{eff}}$ with the new reference dataset is lower, but still within the typical uncertainties of M dwarfs. However, most importantly, the accuracy of the $T_{\mathrm{eff}}$ determinations with the new reference dataset is greater, being in closer agreement with several other methods in literature, as shown later in Sect.~\ref{comparison5}. 

Similarly to \citet{antoniadis-karnavas20}, we examined the parameter determination, model errors and dispersion for stars with spectra of different spectrographs comparing the results to their reference values as defined by the new reference dataset. These results are presented in Table~\ref{upgradenewstarspar}. The predictions of ODUSSEAS had differences varying from 5 to 64 K for $T_{\mathrm{eff}}$ and from 0.00 to 0.05 dex for [Fe/H] compared to the expected values.
We also injected the intrinsic uncertainties of our new reference values to the new training datasets, as in \citet{antoniadis-karnavas20}. These are 99 K for $T_{\mathrm{eff}}$ and 0.17 dex for [Fe/H], as reported by \citet{khata21} and \citet{neves12}, respectively. We created Gaussian distributions on the parameters for each HARPS training dataset by perturbing their values according to their uncertainties. They lead to machine learning model errors varying from 90 to 99 K for $T_{\mathrm{eff}}$ and from 0.11 to 0.13 dex for [Fe/H], as the resolution of the spectra becomes lower. The predictions of the parameters remained within these error ranges compared to the expected reference values. These results take the intrinsic uncertainties into consideration and are presented in Table~\ref{upgradegdstarpar}.

\begin{table*}
\centering
\caption[Determination of parameters, dispersion and model errors applying the new reference scale]{Machine learning (M.L.) results of $T_{\mathrm{eff}}$ and [Fe/H], their dispersion (Disp.), the mean absolute errors (M.A.E.) of the models, and the reference values (Ref.) by applying the new reference scale.}
\begin{tabular}{cccccccccc}
\hline\hline\\
Star & Spec. & Ref. & M.L.  & M.A.E.  & Disp. & Ref. & M.L. & M.A.E. & Disp. \\
 & & $T_{\mathrm{eff}}$ & $T_{\mathrm{eff}}$ & $T_{\mathrm{eff}}$ & $T_{\mathrm{eff}}$ & [Fe/H] & [Fe/H] & [Fe/H] & [Fe/H]\\
 & & [K] & [K] & [K] & [K] & [dex] & [dex] & [dex] & [dex] \\
\hline\\
Gl643 & HARPS & 3243 & 3238 & 65 & 11 & -0.32 & -0.31 & 0.04 & 0.01 \\
Gl846 & UVES & 3810 & 3802 & 65 & 21  & -0.07 & -0.06 & 0.04 & 0.02 \\
Gl514 & CARMENES & 3671 & 3624 & 65 & 33 & -0.15 & -0.15 & 0.04 & 0.03 \\
Gl908 & SOPHIE & 3475 & 3507 & 65 & 53 & -0.40 & -0.36 & 0.04 & 0.03 \\
Gl674 & FEROS & 3409 & 3473 & 65 & 42 & -0.19  & -0.14 & 0.04 & 0.03 \\
\hline\\
\end{tabular}
\label{upgradenewstarspar}
\end{table*}

\section{Application to ESPRESSO data of the highest resolution}
\label{espresso}

After the analysis of the efficiency towards lower resolutions and the successful application to the spectrographs mentioned above, the tool has been also applied successfully on spectra from ESPRESSO\footnote{\url{https://www.eso.org/sci/facilities/paranal/instruments/espresso.html}} \citep{pepe21} spectrograph, that were made available to us. 
%ESPRESSO\footnote{\url{https://www.eso.org/sci/facilities/paranal/instruments/espresso.html}}
This case is special because the high resolution (HR=140000) and ultra-high resolution (UHR=190000) of ESPRESSO spectra are higher than the resolution (R=115000) of the HARPS reference spectra. 
For these very high resolutions of ESPRESSO and HARPS, the results derived by ODUSSEAS are essentially the same when using either directly the original higher-resolution spectra of ESPRESSO that are to be analyzed according to the HARPS reference spectra or by degrading them first to the lower (but still high) resolution of HARPS before being analyzed.
Between the two procedures, the typical differences in the resulting parameters are smaller than the mean errors of our machine learning models.
%Furthermore, the dispersion in the mean values of parameters after the 100 determinations, are similarly low values too.
Therefore, we can analyze the original ESPRESSO spectra, as the process is less time-consuming than analyzing the respective degraded spectra and will still be able to determine their parameters with the same accuracy and precision.
Studies that have used ODUSSEAS tool on ESPRESSO spectra have been already published for the M dwarfs LHS1140 \citep{lillobox20}, L98-59A \citep{demangeon21}, and TOI-3235 \citep{hobson23}, applying the original reference dataset, as well as for TOI-244 \citep{castrogonzalez23}, applying the new reference dataset.

\begin{table*}
\centering
\caption[Determination of parameters, dispersion and model errors after injecting the reference uncertainties to the new training datasets]{Machine learning (M.L.) results of $T_{\mathrm{eff}}$ and [Fe/H] after injecting uncertainties with Gaussian distributions of 99 K and 0.17 dex in the parameters of the new training HARPS datasets, their dispersion (Disp.), the mean absolute errors (M.A.E.) of the models, and the reference values (Ref.) for comparison.}
\begin{tabular}{cccccccccc}
\hline\hline\\
Star & Spec. & Ref. & M.L. & M.A.E. & Disp. & Ref. & M.L. & M.A.E. & Disp. \\
 & & $T_{\mathrm{eff}}$ & $T_{\mathrm{eff}}$ & $T_{\mathrm{eff}}$ & $T_{\mathrm{eff}}$ & [Fe/H] & [Fe/H] & [Fe/H] & [Fe/H]\\
 & & [K] & [K] & [K] & [K] & [dex] & [dex] & [dex] & [dex] \\
\hline\\
Gl643 & HARPS & 3243 & 3227 & 90 & 95 & -0.32  & -0.31 & 0.11 & 0.13 \\
Gl846  & UVES & 3810 & 3794 & 92 & 107 & -0.07  & -0.11 & 0.11 & 0.18 \\
Gl514 & CARMENES & 3671 & 3625 & 95 & 127 & -0.15 & -0.12 & 0.12 & 0.23 \\
Gl908 & SOPHIE & 3475 & 3502 & 97 & 130 & -0.40 & -0.37 & 0.13 & 0.23 \\
Gl674 & FEROS & 3409 & 3481 & 99 & 149 & -0.19  & -0.16 & 0.13 & 0.24 \\
\hline\\
\end{tabular}
\label{upgradegdstarpar}
\end{table*}

\section{Determination of parameters for 82 planet-host stars in SWEET-Cat}
\label{Sweet}

The Catalogue of Stars With ExoplanETs (SWEET-Cat\footnote{\url{https://sweetcat.iastro.pt/}}) contains planet-host stars and their parameters. It was originally introduced in \citep{santos13}. Since then, many more exoplanets have been confirmed, thereby significantly increasing  the number of host stars listed there \citep{sousa21}. 
A crucial step for a comprehensive analysis of the exoplanets is the accurate and precise characterization of their host stars. Stellar parameters derived from high-resolution and high-S/N spectra can provide updated and precise parameters for the discovered planets.
Statistical studies of exoplanets and their host stars require consistent measurements in order to avoid different systematics affecting the results.
SWEET-Cat aims to provide stellar parameters in a consistent and homogeneous way, by analyzing high-quality spectra which are assembled for stars hosting planets. 
%SWEET-Cat\footnote{\url{https://sweetcat.iastro.pt/}}

In this work, we have determined $T_{\mathrm{eff}}$ and [Fe/H] values for 82 stars of SWEET-Cat. 
Among the planet-host stars in the current SWEET-Cat archive, we searched for the stars with $T_{\mathrm{eff}}$ lower than 4000~K and $G$ magnitude brighter than 13.
The ESO archive\footnote{\url{http://archive.eso.org/wdb/wdb/adp/phase3_main/form}} is a main source of reduced spectral public data for SWEET-Cat. We searched the required data in the ESO archive using the "astroquery.eso" module and we downloaded data from the HARPS, ESPRESSO, UVES, and FEROS spectrographs.
In most cases we downloaded more than one spectrum per star to achieve a higher S/N by combining them. 
We discarded spectra with S/N below 20, due to their bad quality. We co-added spectra until we reached a maximum S/N of 2000 for the combined spectrum.
We obtained the final combined spectrum for each star for a given instrument by shifting all spectra to the same wavelength reference before the flux co-addition, using a cross-correlation function with the highest S/N individual spectrum taken as the mask. We used a sigma clipping to remove unusual peaks.
We saved the final spectra in 1D fits-file format.
On top of the ESO archive, we did the same process to collect spectra for more planet-host M dwarfs in SWEET-Cat, using the HARPS-N spectrograph (archived in the IA2\footnote{\url{http://archives.ia2.inaf.it/tng/}}, Italian Center for Astronomical Archive), the SOPHIE\footnote{\url{ http://atlas.obs-hp.fr/sophie/}} spectrograph and the public archive of CARMENES\footnote{\url{http://carmenes.cab.inta-csic.es/gto/jsp/reinersetal2017.jsp}} spectrograph. 
% ESO archive\footnote{\url{http://archive.eso.org/wdb/wdb/adp/phase3_main/form}}
% IA2\footnote{\url{http://archives.ia2.inaf.it/tng/}}
% SOPHIE\footnote{\url{ http://atlas.obs-hp.fr/sophie/}}
% CARMENES\footnote{\url{http://carmenes.cab.inta-csic.es/gto/jsp/reinersetal2017.jsp}}

\begin{figure}
\centering
$\begin{array}{c}
\includegraphics[width= \hsize]{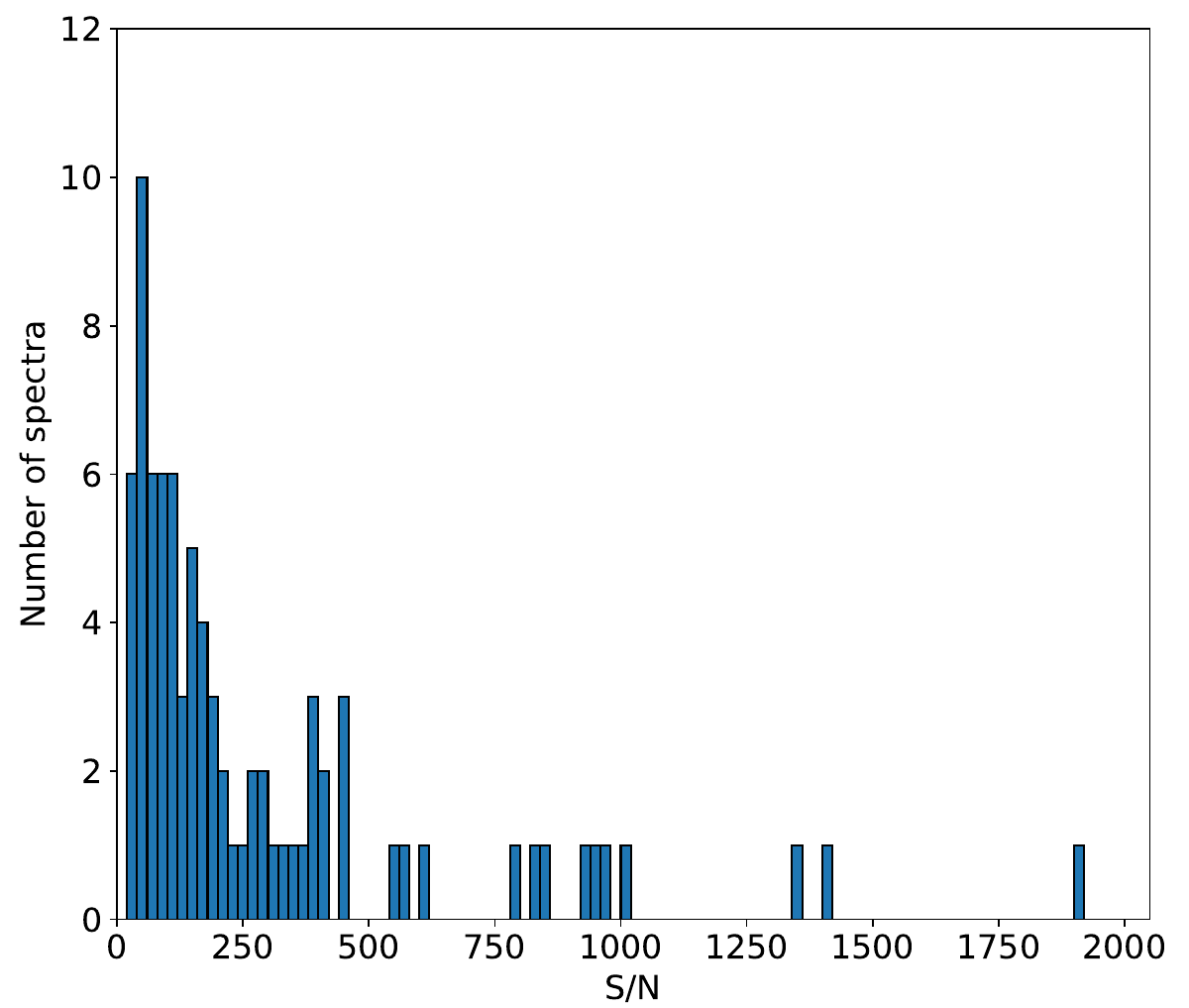} \\
\includegraphics[width= \hsize]{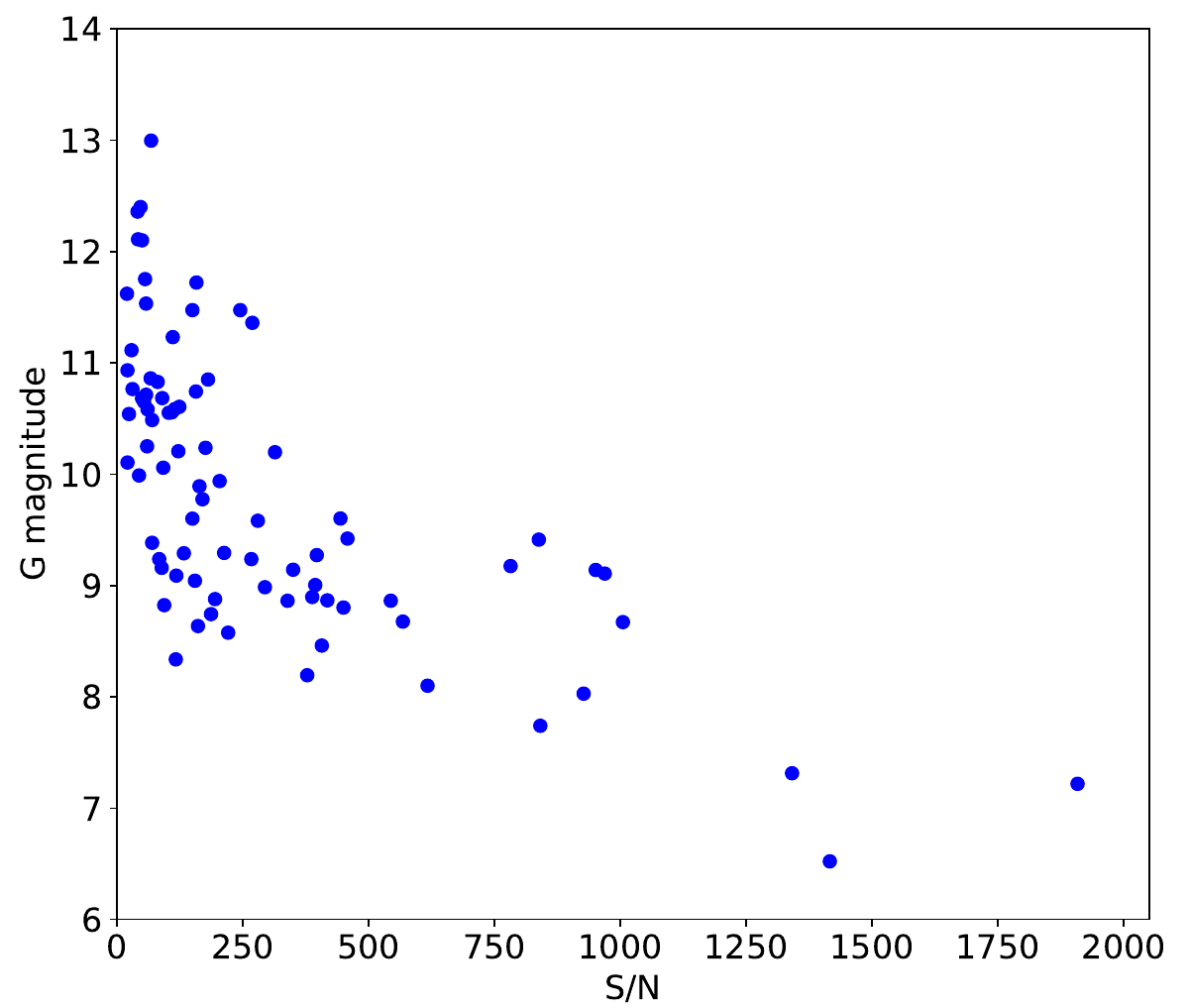}\\
\end{array}$
\caption[Distribution of S/N values and their relation to the $G$ magnitude for spectra of the 82 stars in SWEET-Cat.]{Distribution of S/N values (upper panel) and their relation to the $G$ magnitude (lower panel) for spectra of the 82 stars in SWEET-Cat.}
\label{SNRdistribution} 
\end{figure}

\begin{figure}
\centering
$\begin{array}{c}
\includegraphics[width= \hsize]{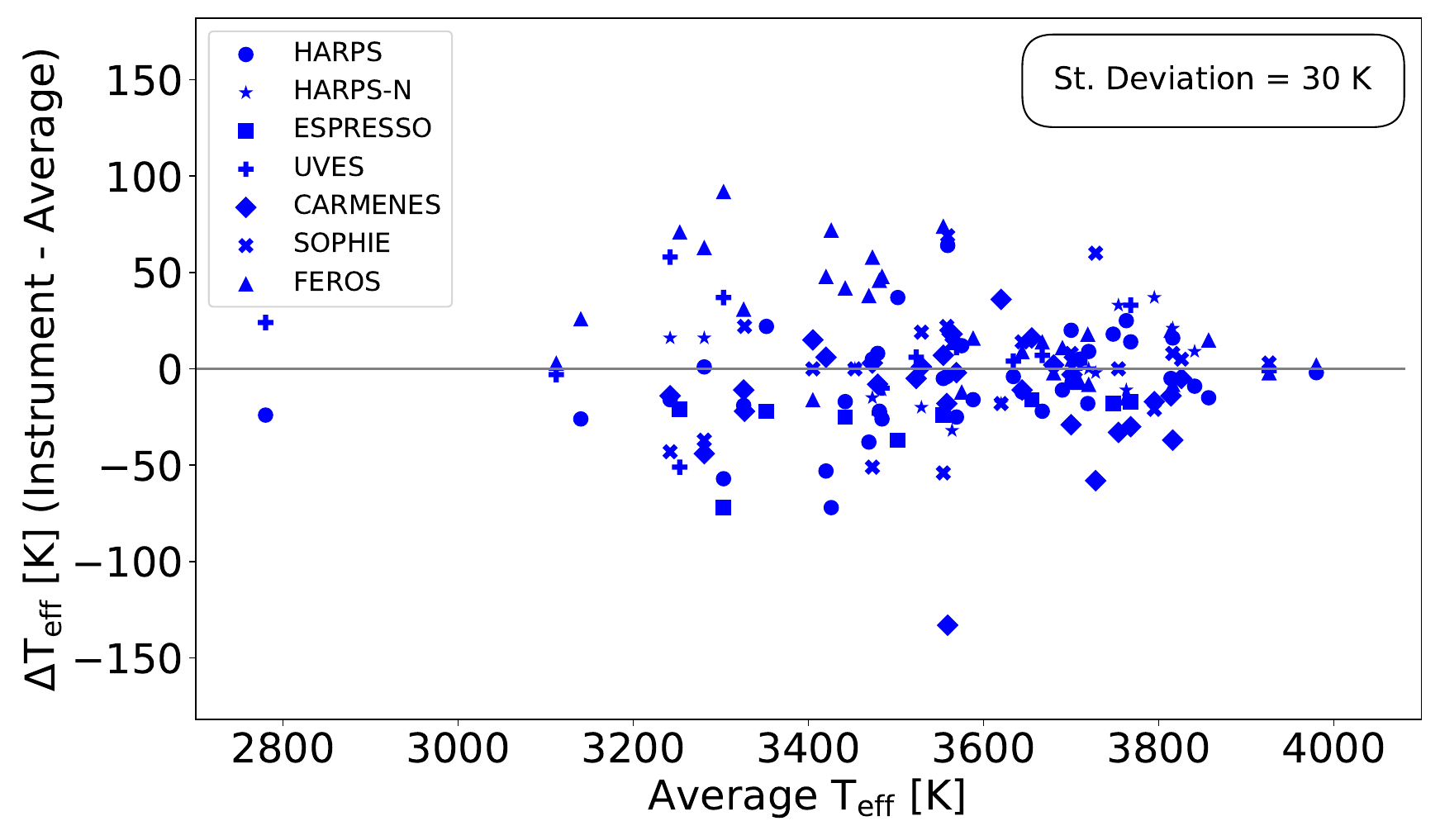} \\
\includegraphics[width= \hsize]{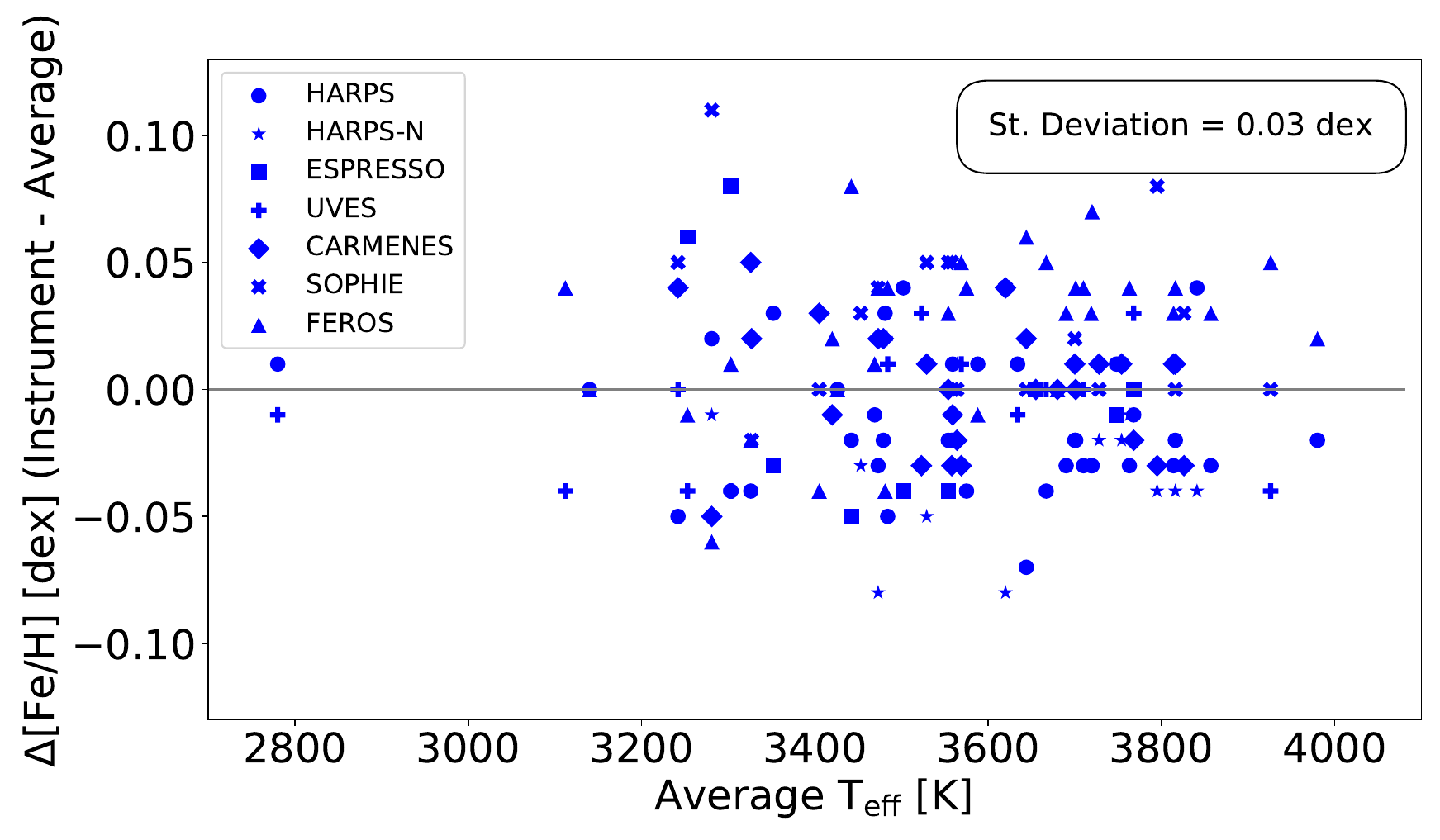}\\
\end{array}$
\caption[Comparison of determinations for stars with spectra from different instruments]{Comparison of determinations for stars with spectra from different instruments. The differences are relative to the average values of $T_{\mathrm{eff}}$ (upper panel) and [Fe/H] (lower panel).}
\label{multiinst} 
\end{figure}

For some stars, we have co-added spectra from more than one instrument. We determined parameters for spectra from all the instruments to check the consistency of ODUSSEAS when using different instrumental data. For the next steps of our work, we selected those parameter values with the lowest estimated errors, instead of the average values, because the average can still be significantly affected by the presence of lower-quality spectra.
The distribution of S/N values for the co-added spectra with the selected parameters and their correlation with the $G$ magnitudes of the stars are presented in Fig.~\ref{SNRdistribution}. The lowest S/N value in our spectra is 20, equal to the minimum value for which ODUSSEAS can provide reliable results \citep{antoniadis-karnavas20}.
The stars and their parameters, determined  with both the new and the old reference datasets, are reported in Table~\ref{Bdet}. 
The total error budgets are reported for each star individually. These are derived automatically when running ODUSSEAS, by adding quadratically the dispersion of the resulting stellar parameters and the maximum errors of the machine learning models at the respective resolution, after having taken into consideration the intrinsic uncertainties of the reference dataset during the machine learning process. 
In Fig.~\ref{multiinst} we show the differences of the derived parameters for spectra of the same stars observed with different instruments. The differences are calculated compared to the average value from the multiple spectra as measured applying the new reference dataset. The values are quite consistent when ODUSSEAS is applied to different instruments. The standard deviation of the parameters, relative to the average values, is 30 K for $T_{\mathrm{eff}}$ and 0.03 dex for [Fe/H]. There are some large individual differences, but they are within the uncertainties reported.

\begin{figure}
\centering
$\begin{array}{c}
\includegraphics[width= \hsize]{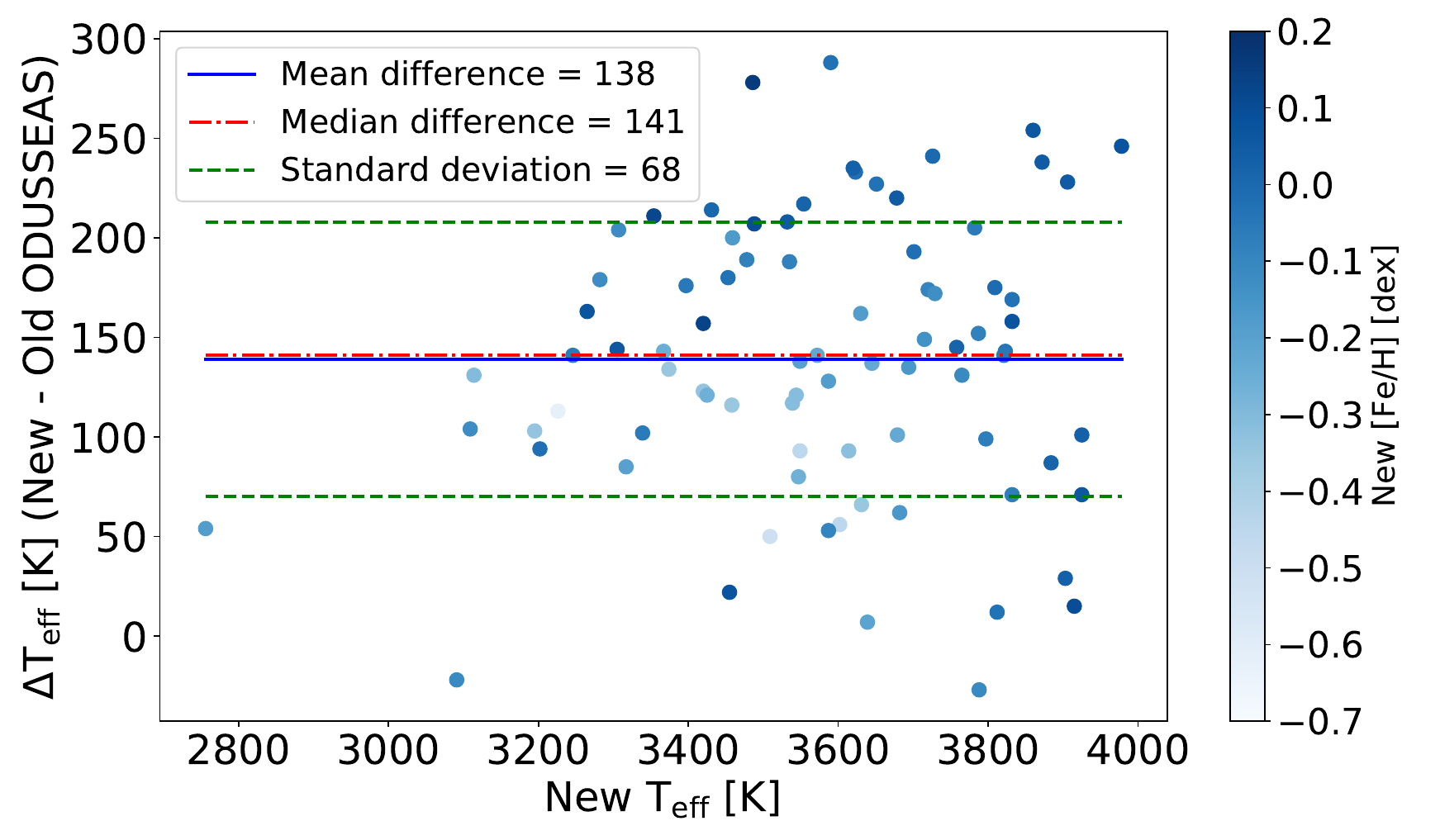} \\
\includegraphics[width= \hsize]{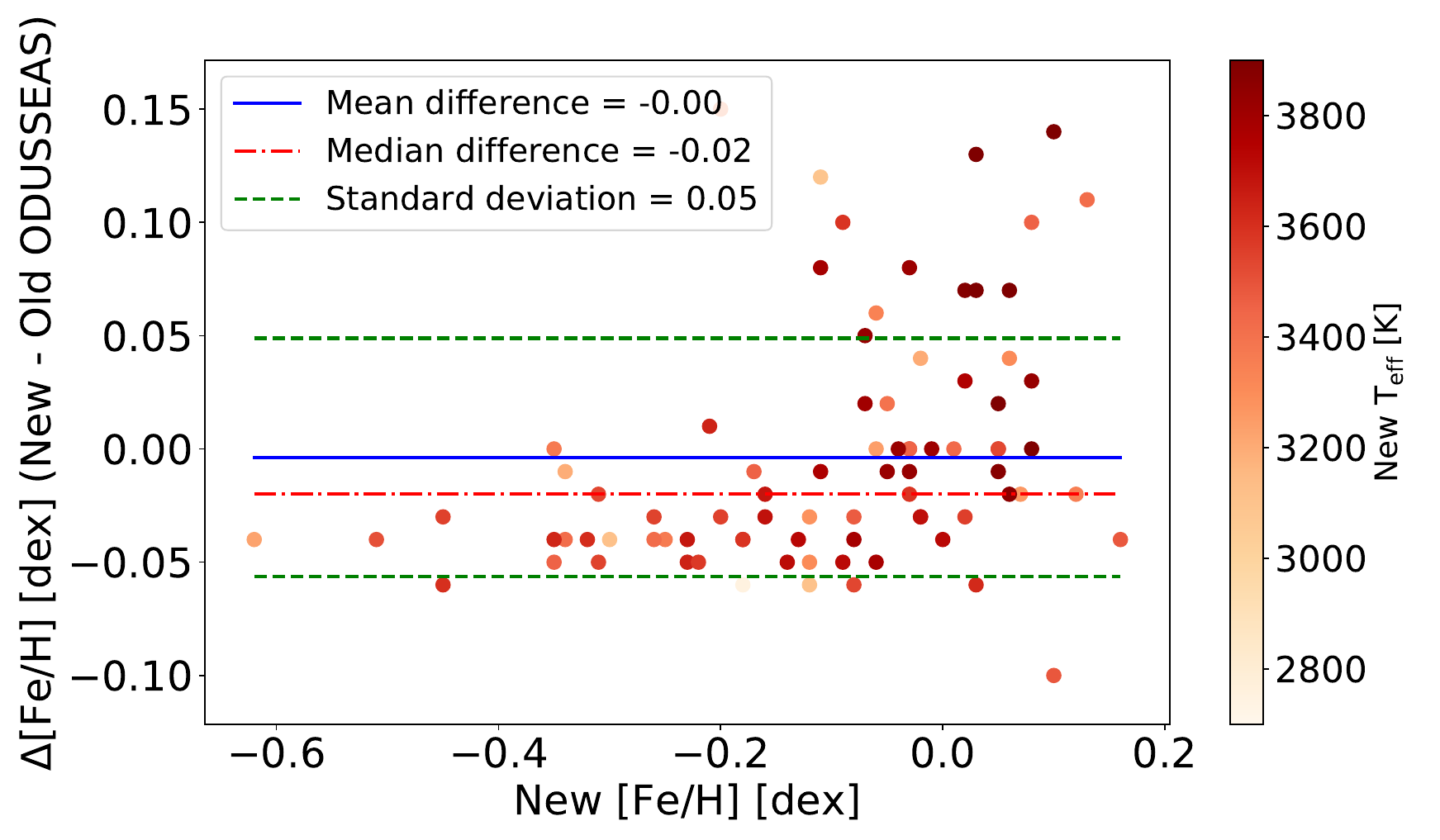}\\
\end{array}$
\caption[Difference of determinations for 82 stars in SWEET-Cat by applying the new and the old reference datasets]{$T_{\mathrm{eff}}$ comparison (upper panel) and [Fe/H] comparison (lower panel) for 82 stars in SWEET-Cat by applying the new and the old reference datasets.}
\label{82comparison} 
\end{figure}

The comparisons of selected determinations for all stars between the applications of the two reference datasets are presented in Fig.~\ref{82comparison}. There is an average difference of 138 K with standard deviation of 68 K between measurements occurring from the new and the old reference dataset. This value of difference is what we expected, based on the differences reported in the comparative study of different methods \citep{passegger22} when we had used the original reference dataset. Thus, the new interferometry-based $T_{\mathrm{eff}}$ scale compensates the underestimation of the order of 140 K that was noticed and explained in Sect.~\ref{intro}. It seems to provide indeed more accurate values compared to the ones derived by the old reference scale, which is based on the IRFM method. 
Regarding the [Fe/H] results in the lower panel, a small trend in differences is noticed for the metal-poor stars, which have slightly lower values with the new reference dataset.
On the other hand, stars with [Fe/H] results close to solar values and above appear to have either higher or lower values with the new reference dataset. 
These differences may occur due to the different samples of reference stars between the two datasets. Another reason may be the updated parallax values that are used to derive the reference [Fe/H] in the new scale, which originate from Gaia eDR3, compared to the parallax values from Hipparcos that were used in the calculation of [Fe/H] by \citet{neves12} in the initial reference dataset. Among the common stars in the two reference datasets, the new reference [Fe/H] scale has an average difference of -0.01 dex with standard deviation of 0.02 dex compared to the initial one. The average difference in our results, applying the two reference datasets, is 0.00 dex with standard deviation of 0.05 dex. 
Having color-coded the datapoints of each panel with the values of the other parameter, looking at the $T_{\mathrm{eff}}$ differences in the upper panel, we notice that the extremely metal-poor stars of the sample tend to have comparatively smaller differences between the two reference scales (from 50 K to 150 K difference), rather than the stars with [Fe/H] close to solar values and above, for which most of the differences are mainly from 150 K up to 300 K. 
In any case, users now have the option to set which reference version is meant to be applied for parameter determinations.

\section{Comparison of results with literature studies}
\label{comparison5}

\begin{figure}
\centering
$\begin{array}{c}
\includegraphics[width= \hsize]{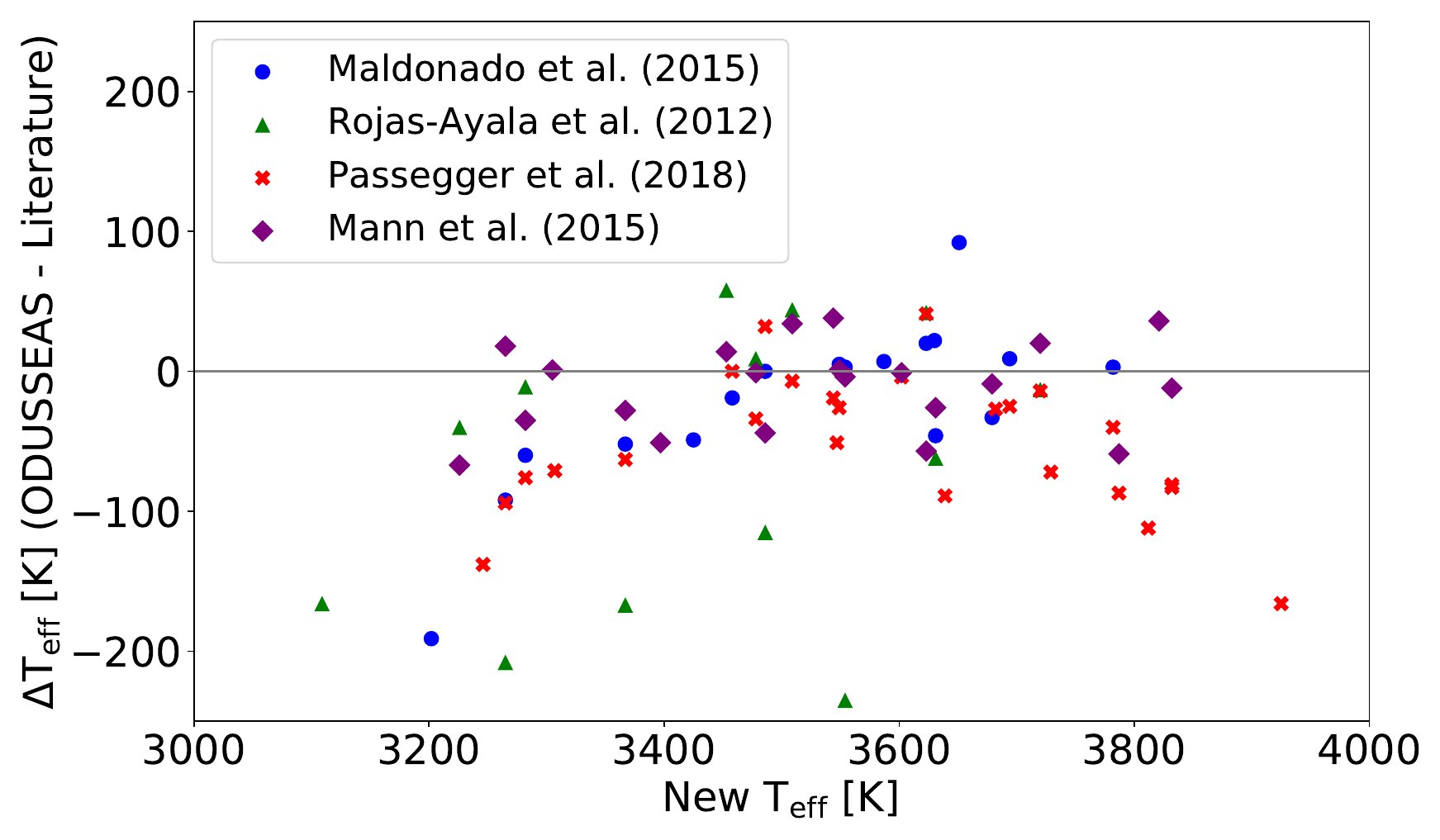} \\
\includegraphics[width= \hsize]{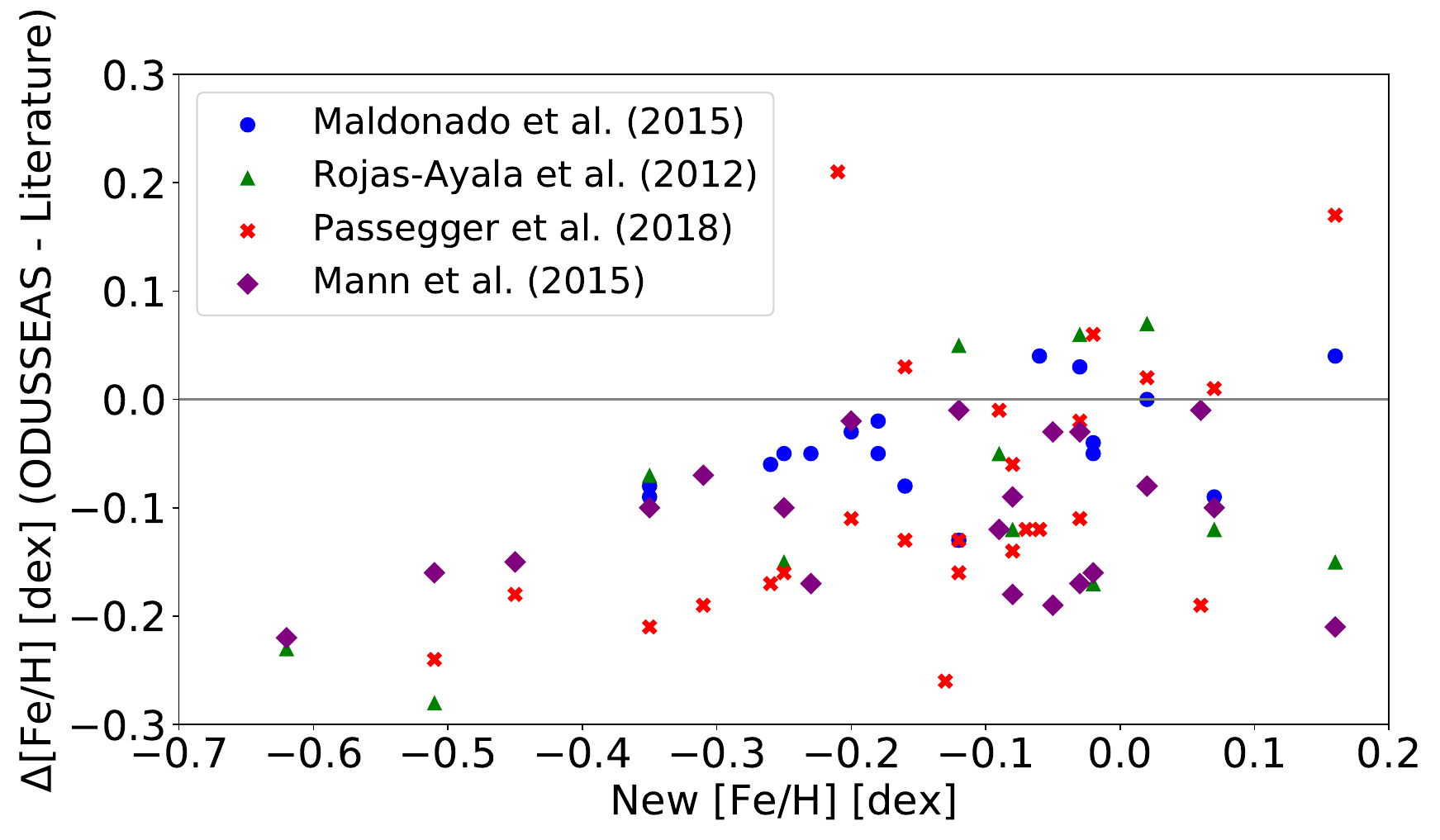}\\
\end{array}$
\caption[Comparison of parameters between our results and studies in the literature.]{$T_{\mathrm{eff}}$ comparison (upper panel) and [Fe/H] comparison (lower panel) between our results and studies in the literature for stars in common.}
\label{4comp} 
\end{figure}

Having determined the parameters for the 82 M dwarfs, we proceeded to perform a comparison of our results (derived by applying the new reference dataset) with studies in the literature, with which we shared a sufficient amount of common stars (at least 15\% of our sample) to ensure that the overall differences are of statistical significance.
Thus, we selected four publications, with at least 13 stars in common each of them, in order to compare the determinations of $T_{\mathrm{eff}}$ and [Fe/H]. 
\citet{maldonado15} and \citet{rojasayala12} derived the stellar parameters applying the concept of pseudo-EWs, while \citet{passegger18} and \citet{mann15} used spectral synthesis with PHOENIX models. 
The differences of parameters for stars in common are presented in Fig.~\ref{4comp}.

\citet{maldonado15} used 112 ratios of pseudo-EWs of spectral features as a temperature diagnostic. The sample of 21 M dwarfs from \citet{boyajian12}, with $T_{\mathrm{eff}}$ obtained from interferometric estimates of their radii, were used as calibrators. 
Combinations of features and ratios of features were used to derive calibrations for the determination of [Fe/H]. They assembled a list of 47 metallicity calibrators with available HARPS spectra, known parallaxes and magnitudes by using the photometric calibration provided by \citet{neves12}.
Regarding $T_{\mathrm{eff}}$, the mean and median differences are -22 K and 0 K, respectively, with a standard deviation of 58 K. There seems to be a positive trend, as for cooler stars our determinations give slightly lower $T_{\mathrm{eff}}$ values and vice versa. The [Fe/H] mean and median differences are small too, reasonably since both methods have used the relation by \citet{neves12} in their respective calibration processes for the determination of [Fe/H]. They are equal to -0.04 and -0.05, respectively, with a standard deviation of 0.05 dex.

\citet{rojasayala12} measured the EWs of the Ca and Na lines, along with the spectral index quantifying the absorption due to H$_2$O opacity (the H$_2$O–K2 index). They set a pseudo-continuum for each feature, which is estimated from a linear fit to the median flux within 0.003 \textmu m wide regions centered on the continuum points. 
Using empirical spectral type standards and synthetic models, they calibrated the H$_2$O–K2 index as an indicator of M dwarf spectral type and $T_{\mathrm{eff}}$. They also presented a relationship that estimates the [Fe/H] and [M/H] metallicities of M dwarfs from their NaI, CaI, and H$_2$O–K2 measurements.
Regarding $T_{\mathrm{eff}}$, the mean and median differences are -66 K and -40 K, respectively, with a standard deviation of 97 K. [Fe/H] mean and median differences are -0.11 and -0.12 dex, respectively, similar to our uncertainty range which varies between 0.11 and 0.14 dex. The standard deviation of the [Fe/H] difference is 0.12 dex.

\citet{passegger18} adapted the method described in \citet{passegger16}, which determined $T_{\mathrm{eff}}$, $\log g$, and [Fe/H] using the grid of PHOENIX model spectra presented by \citet{husser13}. The PHOENIX-ACES model grid they used was especially designed for modeling the spectra of cool dwarfs, because it uses a new equation of state to improve the calculations of molecule formation in cool stellar atmospheres. They applied a $\chi^2$ method to derive the stellar parameters by fitting the models to high-resolution spectroscopic data. 
Regarding $T_{\mathrm{eff}}$, the mean and median differences are -50 K and -45 K, respectively, with a standard deviation of 48 K. The [Fe/H] mean and median differences are -0.09 and -0.12 dex, respectively, similar to the typical uncertainty range of our measurements. The standard deviation of the [Fe/H] difference is 0.12 dex. There seems to be a positive trend, as for stars with low metallicity, our determinations give lower [Fe/H] values, while for higher-metallicity stars, the differences become generally smaller.

\citet{mann15} calculated $T_{\mathrm{eff}}$ by comparing optical spectra with the CFIST suite of the BT-SETTL version of the PHOENIX atmosphere models \citep{allard13}. Metallicities were calculated from the NIR spectra using the empirical relations from \citet{mann13a} for K7–M4.5 dwarfs and from \citet{mann14} for M4.5–M7 dwarfs. These studies provide relations between the EWs of atomic features such as Na and Ca in NIR spectra and the metallicity of the star, calibrated using wide binaries with an FGK primary and an M dwarf companion.
Regarding $T_{\mathrm{eff}}$, the mean and median differences are -11 and -4 K, respectively, with a standard deviation of 31 K. The [Fe/H] mean and median differences are -0.11 and -0.10 dex, respectively, within the typical uncertainty range of our measurements. The standard deviation of the [Fe/H] difference is 0.07 dex.

In all four cases, generally our $T_{\mathrm{eff}}$ values are slightly lower with average differences that are much smaller than the error range of our determinations, which is mostly slightly above 90 K. 
Regarding [Fe/H], the difference with \citet{maldonado15} is considerably small since both methods use, as their calibration scale, the relation by \citet{neves12}; however, this relation was then applied to different set of reference stars and with different values of parallaxes. 
Comparing our [Fe/H] results with the other three methods, we notice that our determinations are on average -0.11 dex lower; that is, exactly as the common error range of our determinations after having taken into account the intrinsic reference uncertainties in our final error estimation.

\begin{figure}
\centering
$\begin{array}{c}
\includegraphics[width= \hsize]{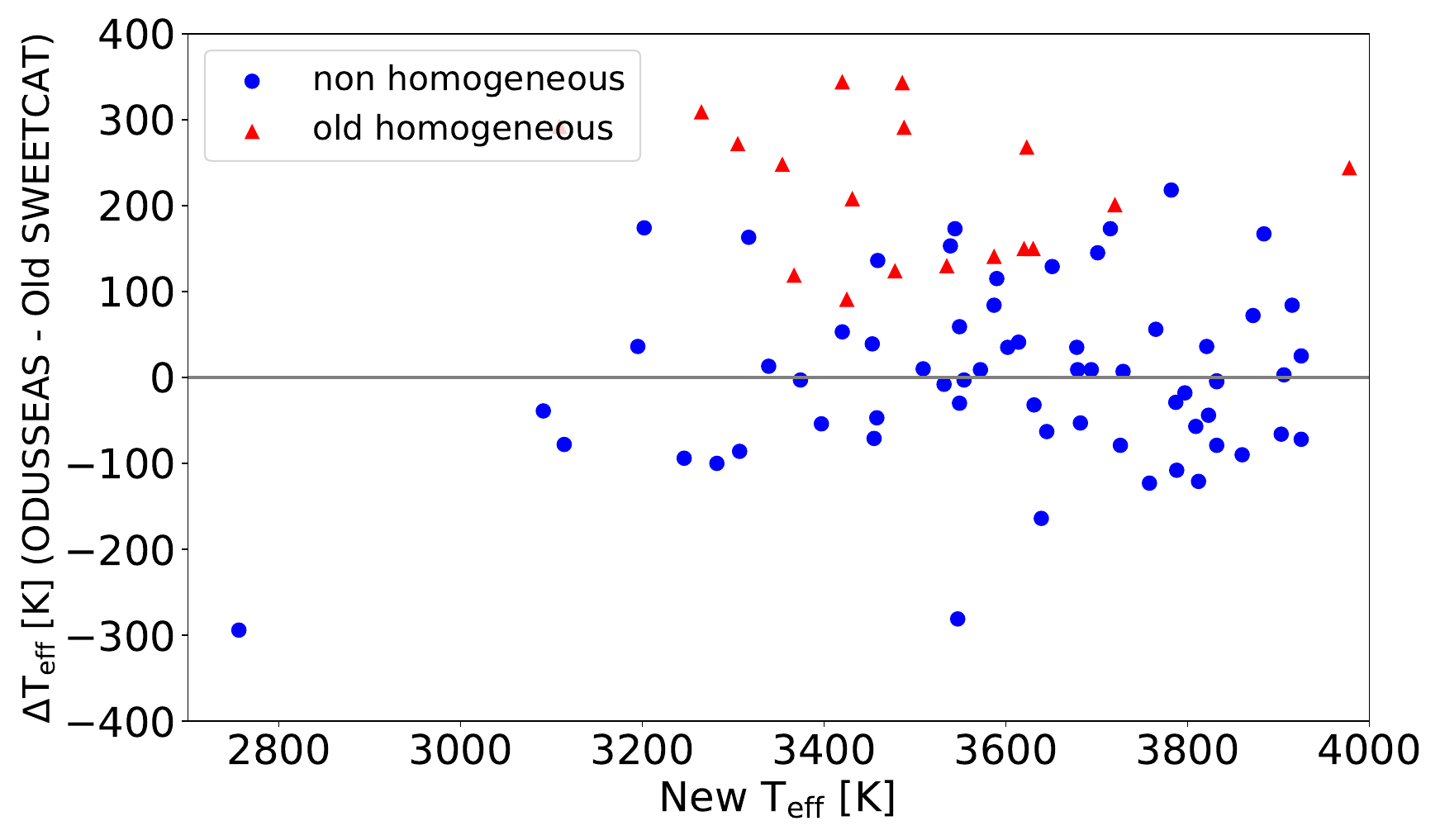} \\
\includegraphics[width= \hsize]{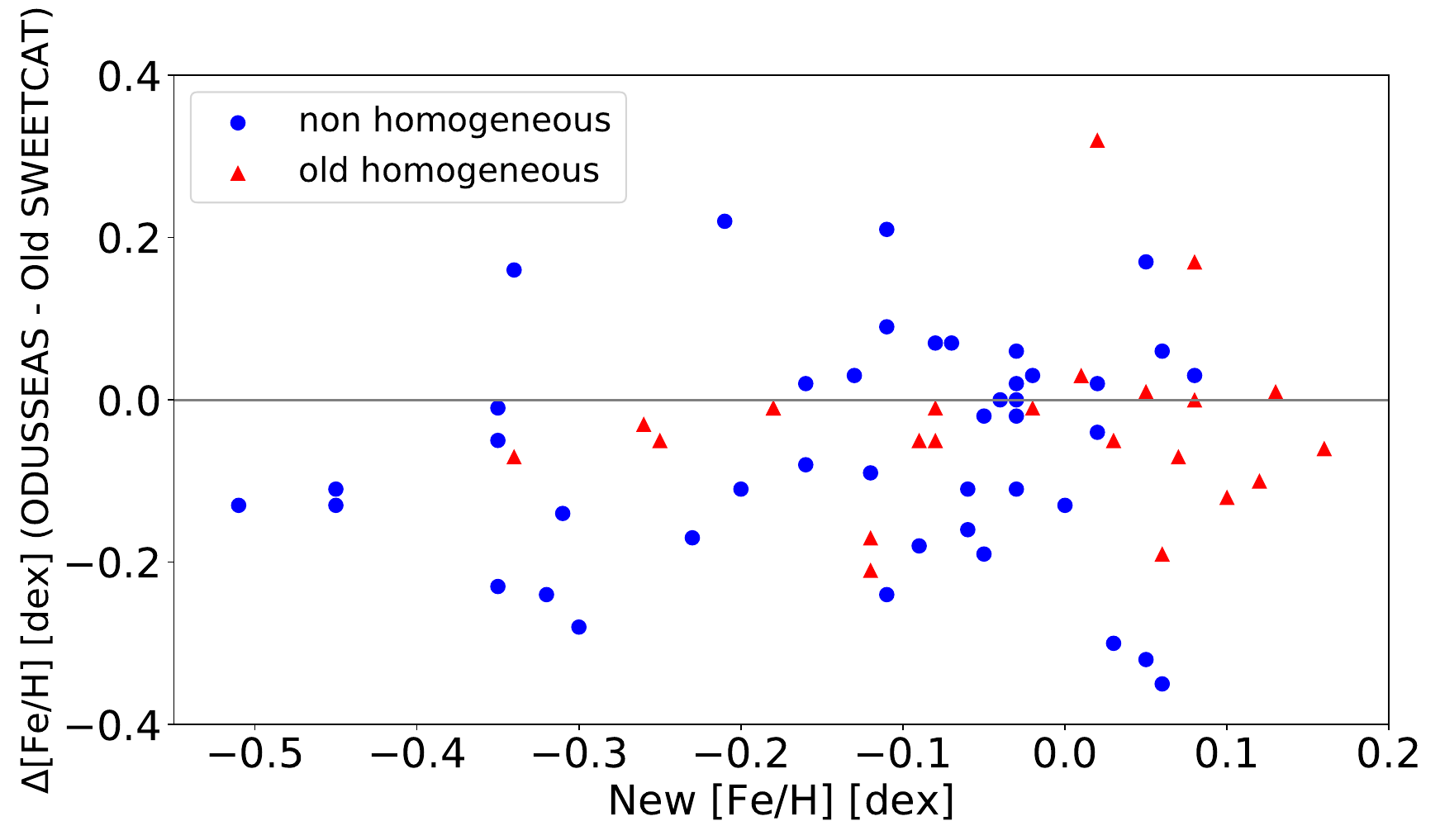}\\
\end{array}$
\caption[Comparison of parameters between our results and previous results reported in SWEET-Cat.]{$T_{\mathrm{eff}}$ comparison (upper panel) and [Fe/H] comparison (lower panel) between our new results and previous results reported in SWEET-Cat.}
\label{Sweetcomp} 
\end{figure}

Furthermore, in Fig.~\ref{Sweetcomp}, we present the differences between our parameters derived with the new reference dataset of ODUSSEAS and the previously reported parameters in SWEET-Cat.
As non-homogeneous parameters, represented with blue circles, we consider those parameters that were listed in SWEET-Cat from different studies in literature following various methods. 
As the old homogeneous parameters, represented with red triangles, we consider the parameters that were derived based on the scales of \citet{casagrande08} and \citet{neves12}, which are the old reference scales of ODUSSEAS as well. The mean and median differences, as well as the standard deviation, regarding the comparisons between our new results and the parameters from all the sources examined above, are summarized in Table~\ref{Comparison_statistics}.

%\begin{table*}
%\centering
%\caption[Comparison statistics]{Mean, median, and standard deviation of differences between our new results applying ODUSSEAS and literature studies.}
%\begin{tabular}{ccccccc}
%\hline\hline\\
%Source & Mean \Delta$T$_{\mathrm{eff}} & Median \Delta$T$_{\mathrm{eff}} & St.D. \Delta$T$_{\mathrm{eff}} & Mean \Delta$[Fe/H]$ & Median \Delta$[Fe/H]$ & St.D. \Delta$[Fe/H]$ \\
% & [K] & [K] & [K] & [dex] & [dex] & [dex] \\
%\hline\\
%Maldonado et al.(2015) & -22 & 0 & 58 & -0.04 & -0.05 & 0.05 \\
%Rojas-Ayala et al. (2012) & -66 & -40 & 97 & -0.11 & -0.12 & 0.12 \\
%Passegger et al. (2018) & -50 & -45 & 48 & -0.09 & -0.12 & 0.12 \\
%Mann et al. (2015) & -11 & -4 & 31 & -0.11 & -0.10 & 0.07 \\
%Non Hom. Sweet-Cat & 1 & -3 & 101 & -0.06 & -0.05 & 0.14 \\
%Old Hom. Sweet-Cat & 218 & 226 & 80 & -0.03 & -0.05 & 0.11 \\
%\hline\\
%\end{tabular}
%\label{Comparison_statistics}
%\end{table*}

\begin{table*}
\centering
\caption[Comparison statistics]{Mean, median, and standard deviation of differences between our new results applying ODUSSEAS and literature studies.}
\begin{tabular}{ccccccc}
\hline\hline\\
Source & Mean $\Delta T_{\mathrm{eff}}$ & Median $\Delta T_{\mathrm{eff}}$ & St.D. $\Delta T_{\mathrm{eff}}$ & Mean $\Delta [Fe/H]$ & Median $\Delta [Fe/H]$ & St.D. $\Delta [Fe/H]$ \\
 & [K] & [K] & [K] & [dex] & [dex] & [dex] \\
\hline\\
Maldonado et al.(2015) & -22 & 0 & 58 & -0.04 & -0.05 & 0.05 \\
Rojas-Ayala et al. (2012) & -66 & -40 & 97 & -0.11 & -0.12 & 0.12 \\
Passegger et al. (2018) & -50 & -45 & 48 & -0.09 & -0.12 & 0.12 \\
Mann et al. (2015) & -11 & -4 & 31 & -0.11 & -0.10 & 0.07 \\
Non Hom. Sweet-Cat & 1 & -3 & 101 & -0.06 & -0.05 & 0.14 \\
Old Hom. Sweet-Cat & 218 & 226 & 80 & -0.03 & -0.05 & 0.11 \\
\hline\\
\end{tabular}
\label{Comparison_statistics}
\end{table*}

\section{Planetary mass-stellar metallicity correlation}
\label{planets}

Theory of planet formation suggests correlation between the presence of giant planets and the high metallicity of their host stars \citep{laughlin00, gonzalez01}.
\citet{santos03, santos04} confirmed that the frequency of planets is a rising function of the stellar metallicity, at least for stars with [Fe/H] > 0.
\citet{fischer05} also concluded that above solar metallicity, there is a smooth and rapid rise in the fraction of stars with planets. Furthermore, in that study, high stellar metallicity appears to be correlated with the presence of multiple-planet systems and with the total detected planet mass. 
The first planets to be found were massive ones and giants, because of the observational biases from the planet detection methods. Later, the correlation was tested for lower-mass planets, where the correlation was not observed \citep{udry06, sousa08, sousa11, buchhave12, wangfischer15}.
Observational works for M-dwarf samples specifically have been in agreement with the correlation of giant-planet occurrence with high stellar metallicity \citep{bonfils07, johnsonapps09, schlaufmanlaughlin10, rojasayala12, terrien12, neves13, maldonado20}. 

\begin{figure}
\centering
$\begin{array}{c}
\includegraphics[width= \hsize]{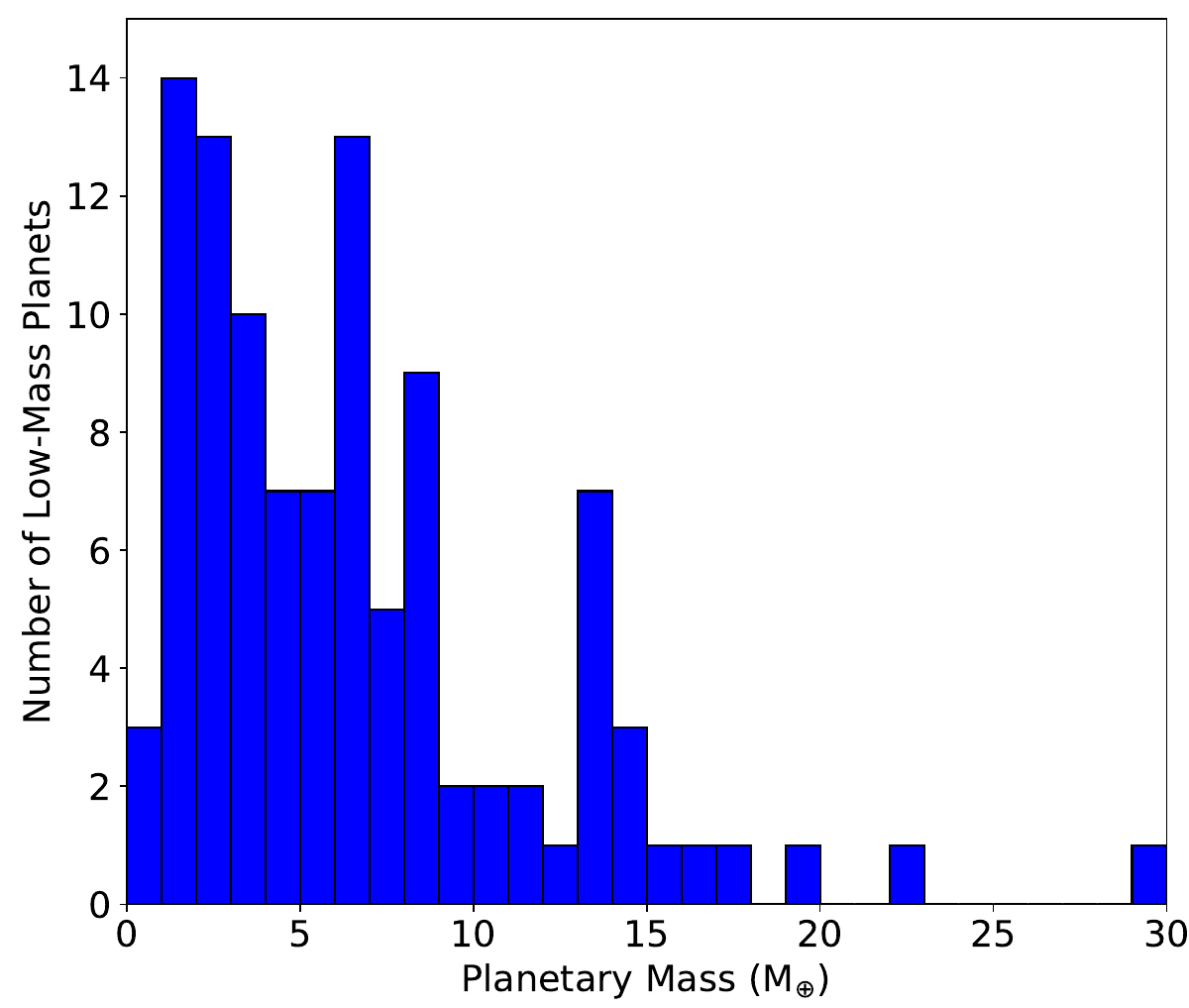} \\
\includegraphics[width= \hsize]{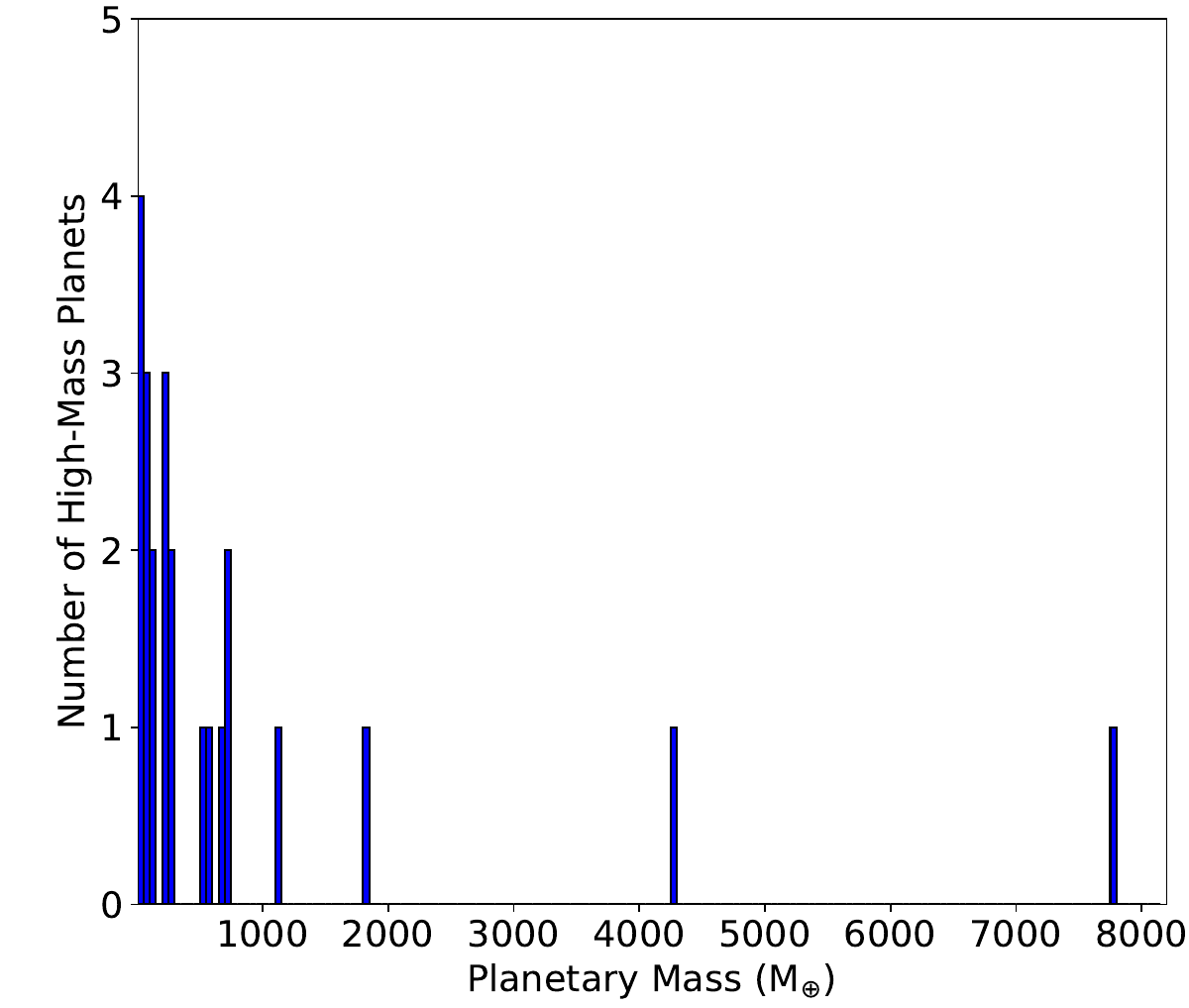} \\
\end{array}$
\caption[Distribution of planetary masses]{Distribution of mass for the low-mass planets (upper panel). Distribution of mass for the high-mass planets (lower panel).}
\label{planetmasses} 
\end{figure}

\begin{figure}
\centering
\includegraphics[width= 0.5\textwidth]{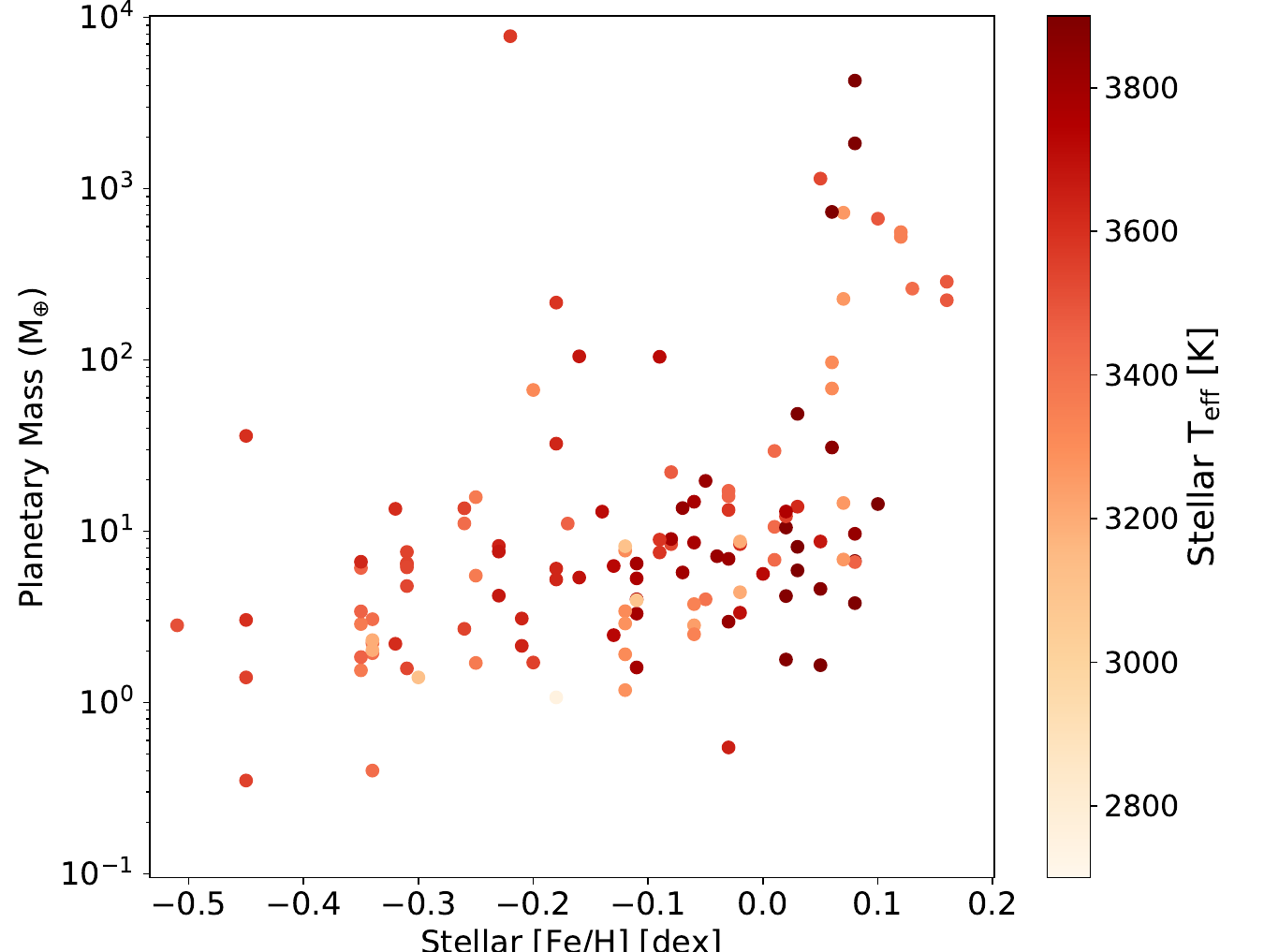} \\
\caption[Distribution of planetary masses against stellar metallicities]{Distribution in logarithmic scale of planetary masses against the host stellar metallicities.}
\label{allmasses} 
\end{figure}

\begin{figure*}
\centering
$\begin{array}{cc}
\includegraphics[width=0.5\textwidth]{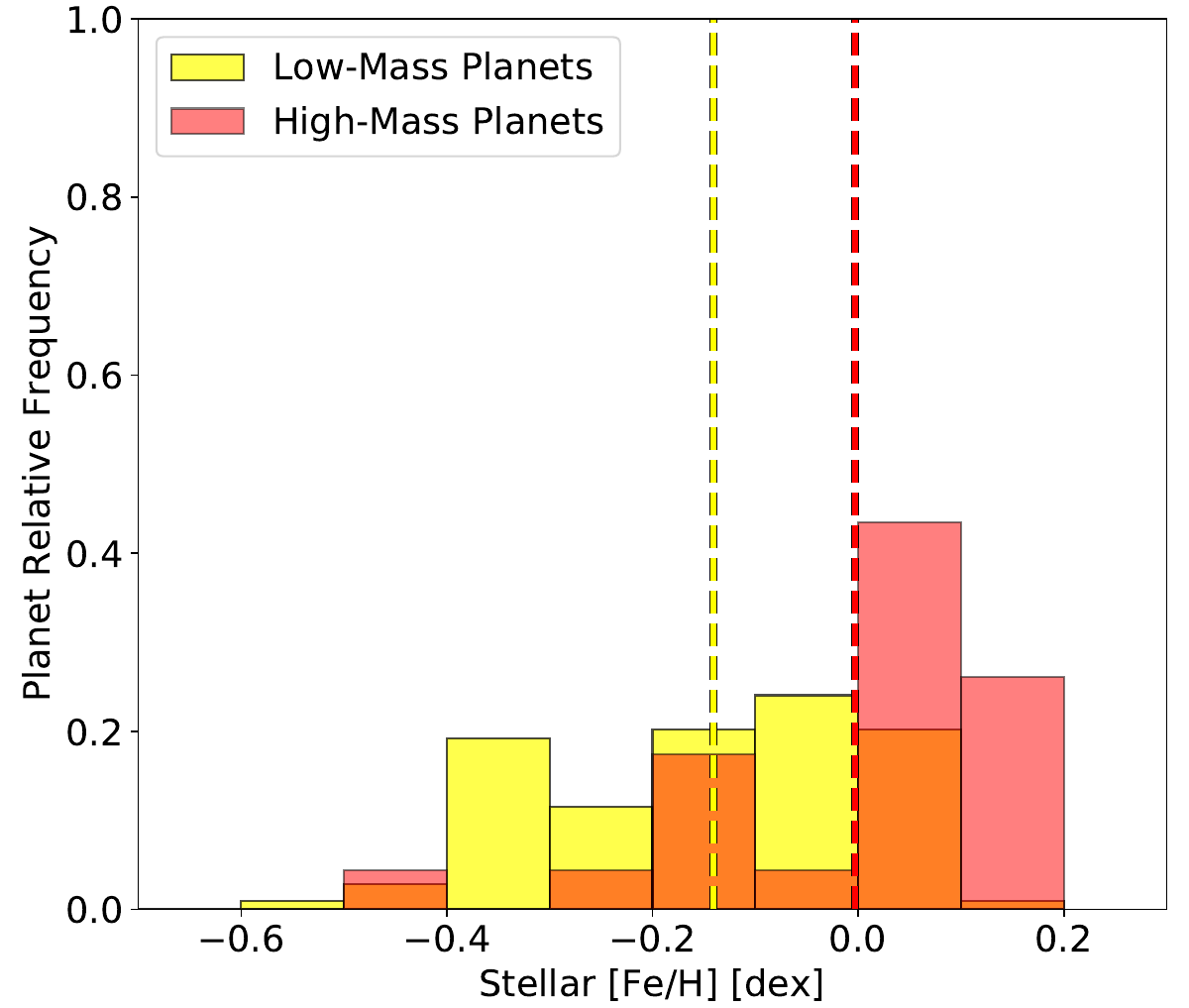} &
\includegraphics[width=0.5\textwidth]{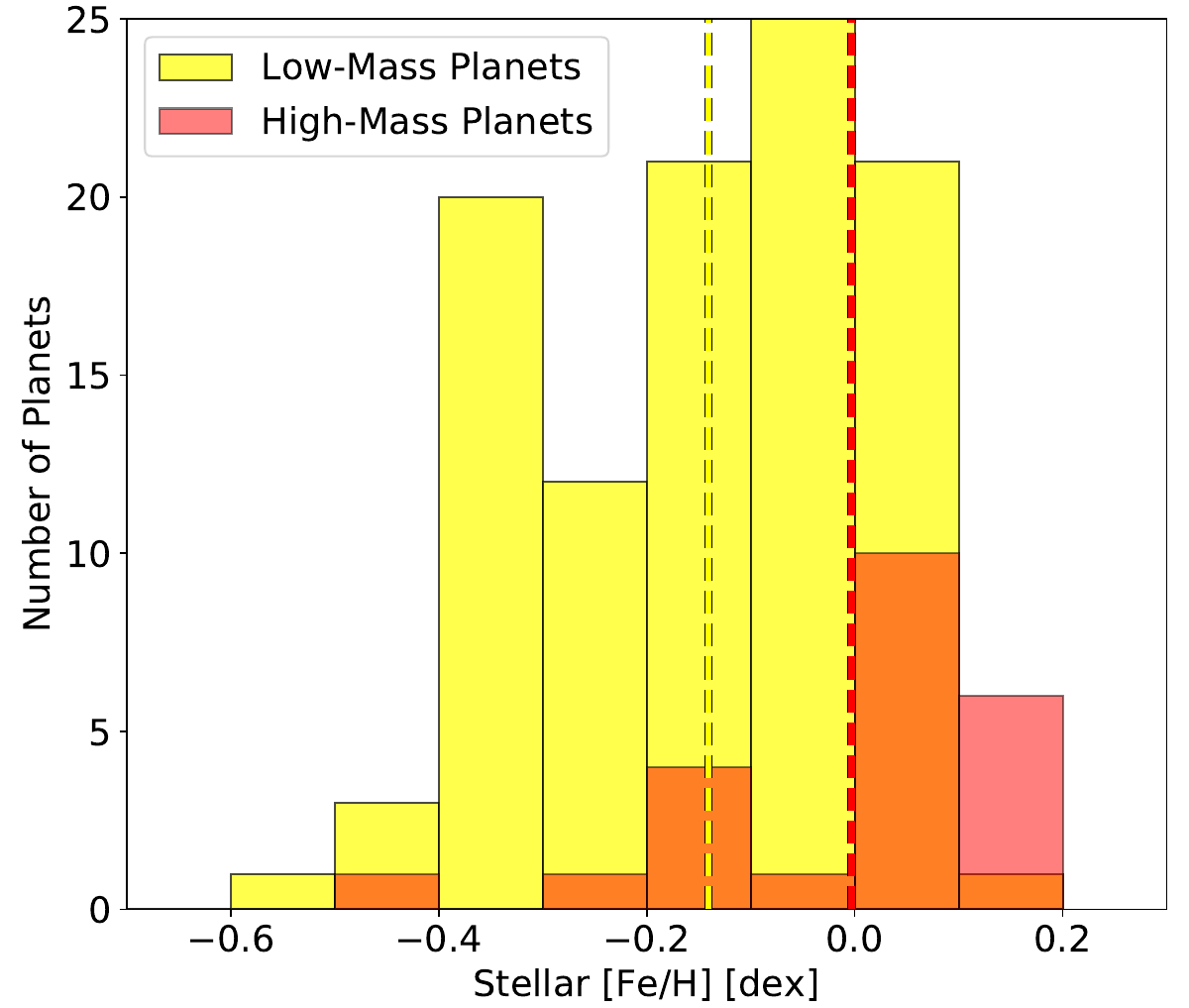} \\
\includegraphics[width=0.5\textwidth]{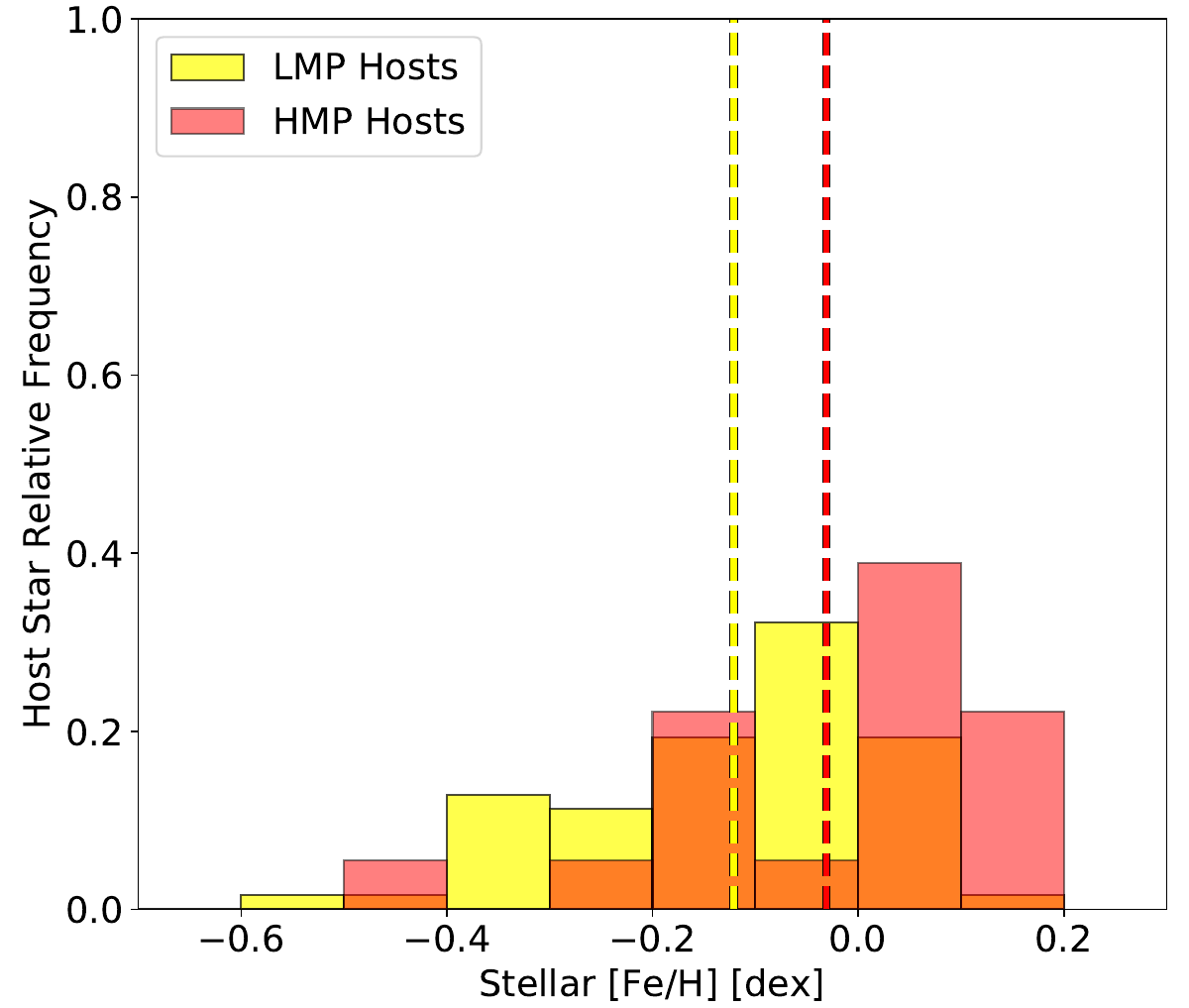} & 
\includegraphics[width=0.5\textwidth]{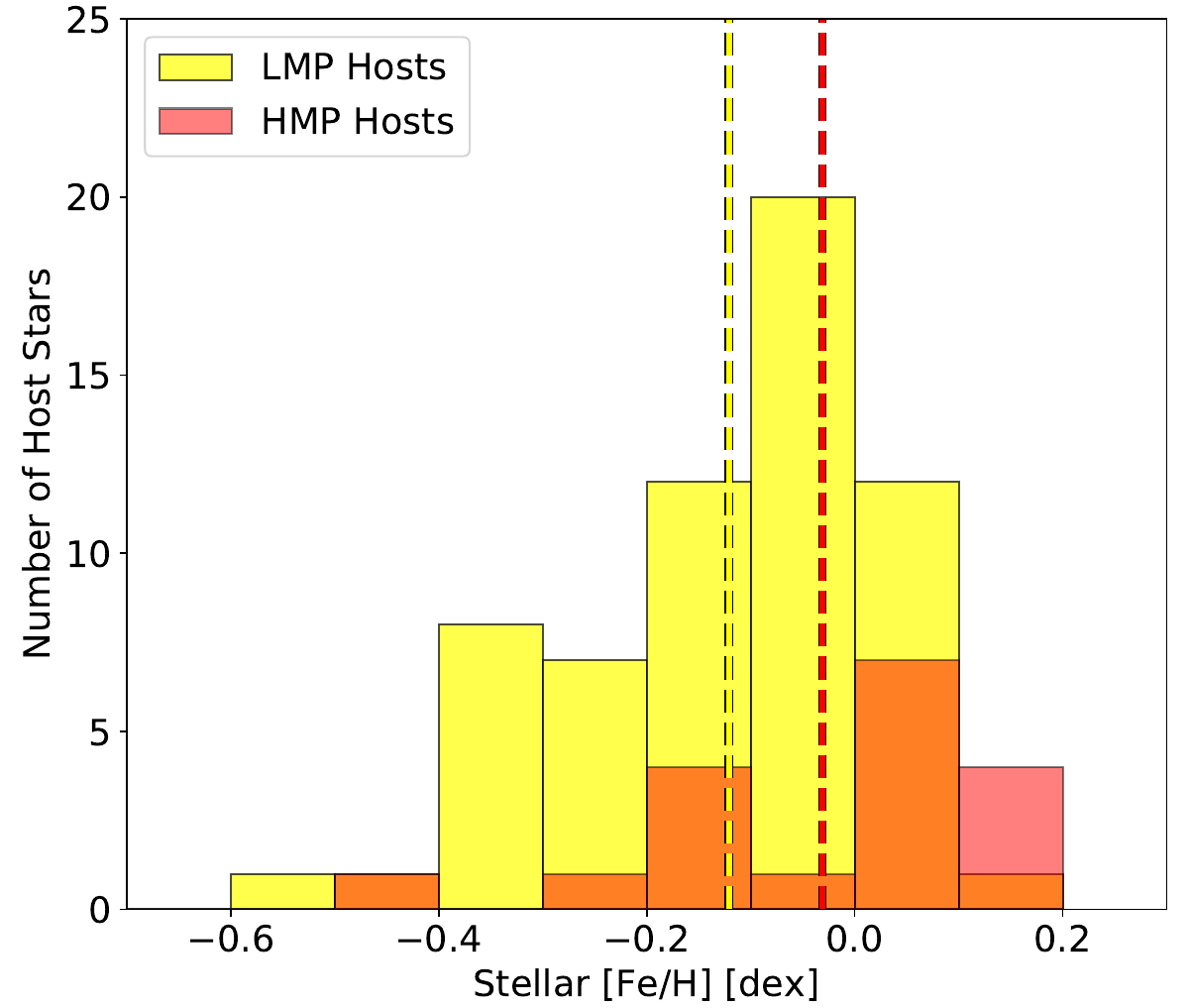}  \\

\end{array}$
\caption[Planetary mass - stellar metallicity correlations]{Planet relative frequency distributions for the LMP and HMP populations (top-left), absolute number of planets (top-right), host star relative frequency for hosting LMP or HMP (bottom-left), and host star absolute numbers (bottom-right). The dashed lines show the mean value of stellar metallicity for each population.}
\label{planetrelation} 
\end{figure*}

% Archive\footnote{\url{https://exoplanetarchive.ipac.caltech.edu/}}
We used our planet-host M dwarfs from SWEET-Cat to examine the correlation between planetary mass and stellar metallicity. The masses of exoplanets were obtained from NASA Exoplanet Archive\footnote{\url{https://exoplanetarchive.ipac.caltech.edu/}}. 
Altogether, 80 stars were included in our analysis: those measured for SWEET-Cat, apart from Gl699 and HIP36985 (which are not listed in the NASA Exoplanet Archive). The total number of planets orbiting those stars is 127. The low-mass planets (LMP: lower than 30~\(\textup{M}_\oplus\)) amount to 104, while there are 23 high-mass planets (HMP: greater than 30~\(\textup{M}_\oplus\)). We chose this division because the location of the gap in the distribution of planetary mass, presented in \citet{mayor11}, is 30~\(\textup{M}_\oplus\). Thus, we kept this threshold so that we could remain consistent with similar works in M dwarfs too, such as \citet{neves13, maldonado20}. For systems with multiple planets, the planet with the highest mass is considered in order to characterize the star as LMP host (LMPH) or HMP host (HMPH). Our stellar sample consists of 62 LMPHs and 18 HMPHs. In Fig.~\ref{planetmasses}, the mass distributions for the low-mass planets and the high-mass planets are presented. 
In Fig.~\ref{allmasses} all the planetary masses in logarithmic scale are presented against the stellar host metallicities.
In Fig.~\ref{planetrelation}, the left panels show the relative frequencies of planets (upper left) and planet-host stars (lower left) against the metallicity distribution of stars as derived by the new version of ODUSSEAS, separated by the two planet populations. As the relative frequency, we consider the way the planets and the host stars are distributed based on the metallicity inside their own population defined by the mass division, respectively. For example, in these panels, we see that the low-mass (yellow) values of relative frequencies all sum up to 1 and the high-mass (red) values all sum up to 1 accordingly. The right panels present the absolute numbers of planets (upper right) and stars (lower right).   

We notice a correlation of high-mass planets increasing at high stellar metallicity. In order to compare the distributions, we applied Kolmogorov-Smirnov tests for low-mass planets versus high-mass planets and LMP hosts versus HMP hosts. The closer the K-S statistic number is to 0, the more likely it is that the two samples are drawn from the same distribution; whereas the closer the value is to 1, the more significant is the difference of the two samples. The p-value refers to the probability of the distributions being drawn by chance. If the p-value is very close to 0 the distributions are statistically different, while for higher p-values the distributions are similar. For the low-mass planet (LMP) versus high-mass planet (HMP) comparison we get a K-S statistic = 0.57 and a p-value = 3.09·$10^{-6}$, while for the low-mass planet host (LMPH) versus the high-mass planet host (HMPH) comparison we get a K-S statistic = 0.48 and a p-value = 1.68·$10^{-3}$. 
These values imply that the difference is statistically significant between the samples of low-mass planets and high-mass planets in relation to metallicity of the host stars. 

Moreover, in order to examine the planetary-mass correlation regarding more stellar characteristics apart from the metallicity, we created similar distributions as before, having divided the host stars in three categories based on their $T_{\mathrm{eff}}$, as indicator of their mass. 
In the same way, we applied Kolmogorov–Smirnov tests for each group of distributions. 
All the results are presented in Table~\ref{KStesttemperatures}. 
The numbers of planets and stars in total, as well as for each group separately, are presented in parenthesis next to each group of objects.
Also, we report the average metallicity and its standard deviation for each group.

Regarding the separate groups based on $T_{\mathrm{eff}}$, we notice a significant difference between the distributions of high-mass and low-mass planets around the cooler stars. In this case, the K-S statistic has the highest value of all, equal to 0.77, and the p-value is low enough, equal to 0.001; this is one order of magnitude lower than the cases of planets around intermediate and hotter stars, which have low p-values too, however (equal to 0.037 and 0.046, respectively). 
This clear difference in the cooler stellar range may be the outcome of a trend (as can be noticed in the lower panel of Fig.~\ref{82comparison})  indicating that the cooler metal-poor stars are measured with our new reference dataset more consistently (at least they have consistent differences from their values measured with the old reference dataset) compared to hotter metal-rich stars. The latter have more scattered differences, when comparing their values to those derived by the old reference dataset. Thus, potential systematic uncertainties in the [Fe/H] values attributed to hotter stars may not make possible for equally clear differences to appear in the distributions of their planets according to stellar [Fe/H].
Another explanation could be that, in order to form high-mass planets at so cool stars (i.e., low masses), very high metallicity is needed to compensate.

From the comparison of distributions for our overall sample of M dwarfs and planets, the known correlation between the presence of giant planets and the high metallicity of their host stars appears to be valid in the case of M-dwarf systems as well. 
This result supports the core-accretion scenario \citep{pollack96} against the disk-instability model \citep{boss00}, as being the main mechanism of giant-planet formation around M dwarfs as well.
According to \citet{boss02}, the formation of a giant planet as a result of disk instabilities is almost independent of the metallicity.
On the other hand, the core-accretion scenario suggests that the higher the metallicity (and thus the dust density of the disk), the higher might be the probability of forming a core (and eventually massive core) before the disk dissipates \citep{pollack96, kokubo02}.
A bigger sample of M-dwarf planet hosts that have spectroscopic data available would strengthen our conclusions.

\begin{table*}
\centering
\caption[Kolmogorov–Smirnov tests comparing the LMP vs HMP and LMPH vs HMPH distributions]{Kolmogorov–Smirnov tests comparing the LMP vs. HMP and LMPH vs. HMPH distributions for the whole sample and for different groups of stellar $T_{\mathrm{eff}}$.}
\begin{tabular}{ccccc}
\hline\hline\\
Samples & $T_{\mathrm{eff}}$ range & K-S statistic & K-S p-value & Mean [Fe/H] (St.D.) \\
\hline\\
LMP (102) vs HMP (23) & all & 0.57 & 3.09·$10^{-6}$ & -0.14 (0.14) vs 0.00 (0.15) \\
LMPH (60) vs HMPH (18) & all & 0.48 & 1.68·$10^{-3}$ & -0.12 (0.14) vs -0.03 (0.16) \\
LMP (24) vs HMP (7) & $T_{\mathrm{eff}}$ < 3400 & 0.77 & 1.01·$10^{-3}$ & -0.15 (0.12) vs 0.04 (0.10) \\
LMPH (13) vs HMPH (4) & $T_{\mathrm{eff}}$ < 3400 & 0.75 & 3.78·$10^{-2}$ & -0.16 (0.11) vs 0.01 (0.12) \\
LMP (49) vs HMP (10) & 3400 < $T_{\mathrm{eff}}$ < 3700 & 0.46 & 3.71·$10^{-2}$ & -0.20 (0.15) vs -0.06 (0.19) \\
LMPH (28) vs HMPH (9) & 3400 < $T_{\mathrm{eff}}$ < 3700 & 0.37 & 2.41·$10^{-1}$ & -0.17 (0.15) vs -0.08 (0.19) \\
LMP (29) vs HMP (6) & $T_{\mathrm{eff}}$ > 3700 & 0.58 & 4.66·$10^{-2}$ & -0.03 (0.07) vs 0.03 (0.06) \\
LMPH (19) vs HMPH (5) & $T_{\mathrm{eff}}$ > 3700 & 0.56 & 1.11·$10^{-1}$ & -0.03 (0.07) vs 0.03 (0.06) \\

\hline\\
\end{tabular}
\tablefoot{In parentheses, we report the numbers of the objects for each case among the samples. We also report the mean metallicity of each sample, with the standard deviation in parentheses, respectively.}
\label{KStesttemperatures}
\end{table*}

\section{Summary}
\label{sum}

We present the upgraded version of our machine learning tool ODUSSEAS for the derivation of $T_{\mathrm{eff}}$ and [Fe/H] in M dwarf stars, with spectra at wavelengths ranging between 530 and 690 nm.
We explain the tests we performed with an aim to examine the accuracy and precision using spectra with resolutions that vary from 190000 down to 48000. We show the results we obtained for M dwarfs in SWEET-Cat and their comparison with the old version of the tool, as well as with results from other studies in the literature.

Our new version of the tool seems to be more accurate, as its results are in agreement with several literature values within the uncertainty range. It operates with high machine learning scores of around 0.90 and achieves predictions with mean absolute errors of $\sim$65 K for $T_{\mathrm{eff}}$ and $\sim$0.04 dex for [Fe/H].
Taking into consideration the intrinsic uncertainties of the reference parameters, our models have maximum uncertainties of $\sim$99 K for $T_{\mathrm{eff}}$ and $\sim$0.13 dex for [Fe/H], which are within the typical uncertainties for M dwarfs.
Our parameters for the spectra of the same stars from different instruments have consistent values with standard deviation of $\sim$30 K and $\sim$0.03 dex from their average values. 

We also examined the correlation between planetary mass and stellar metallicity of our sample, both as a whole and divided in groups based on $T_{\mathrm{eff}}$ of the stars. 
The presence of giant planets mostly around host stars of high metallicity appears evident for M-dwarf systems as well.

ODUSSEAS is valid for M dwarfs in the intervals 2700 to 4000 K for $T_{\mathrm{eff}}$ and -0.82 to 0.24 dex for [Fe/H], except from very active or young stars. The spectra ought to have a S/N of above 20 for optimal predictions. The tool can be used by downloading the files from its GitHub page\footnote{\url{https://github.com/AlexandrosAntoniadis/ODUSSEAS}}, after reading the README instructions for clarifying the technical details and requirements.

\begin{acknowledgements}

This work was supported by FCT - Fundação para a Ciência e a Tecnologia through national funds and by FEDER through COMPETE2020 - Programa Operacional Competitividade e Internacionalização by these grants: UIDB/04434/2020 and UIDP/04434/2020; 2022.04416.PTDC. 
A.A.K., S.G.S. and E.D.M. acknowledge the support from FCT in the form of the
exploratory projects with references IF/00028/2014/CP1215/CT0002, IF/00849/2015/CP1273/CT0003.
S.G.S. further acknowledges the support from FCT through Investigador FCT contract nr. CEECIND/00826/2018 and POPH/FSE (EC).
E.D.M. further acknowledges the support from FCT through the Stimulus FCT
contract 2021.01294.CEECIND.
Funded/Co-funded by the European Union (ERC, FIERCE, 101052347). Views and opinions expressed are, however, those of the authors only and do not necessarily reflect those of the European Union or the European Research Council. Neither the European Union nor the granting authority can be held responsible for them.

\end{acknowledgements}

\bibliographystyle{aa}

%\bibliography{}

\clearpage

\begin{appendix} 
\section{The new reference scale}
\label{A}

Table~\ref{upgraderefparam} is the new reference dataset that was added in the upgraded version of the tool. $T_{\mathrm{eff}}$ values of 43 stars were derived by \citet{khata21} and the values of the last 4 stars at the bottom were derived by \citet{rabus19}. [Fe/H] values were derived by applying the relation from \citet{neves12} with updated parallaxes. The intrinsic uncertainties of reference $T_{\mathrm{eff}}$ and [Fe/H] are reported to be 99 K and 0.17 dex, respectively.

\onecolumn
\begin{longtable}{ccc}
\caption[The new reference dataset of the upgraded ODUSSEAS version]{New reference dataset of the upgraded ODUSSEAS version.} \\
\hline\hline\\
Star & [Fe/H] & $T_{\mathrm{eff}}$  \\
 & [dex] & [K] \\
\hline \\
\endfirsthead
\caption{continued}\\
%\\ \hline 
Star & [Fe/H] & $T_{\mathrm{eff}}$  \\
 & [dex] & [K] \\

\endhead
\hline
\endfoot

Gl54.1 &  -0.42 & 2900 \\
Gl87 &  -0.31 & 3604 \\
Gl105B &  -0.14 & 3251 \\
Gl176 &  0.00 & 3590 \\
Gl179 &  0.13 & 3382 \\
Gl203 &  -0.33 & 3227 \\
Gl205 &  0.10 & 3813 \\
Gl213 &  -0.24 & 3066 \\
Gl229 &  -0.07 & 3796 \\
Gl273 &  -0.07 & 3256 \\
Gl299 &  -0.65 & 3246 \\
Gl300 &  0.09 & 3106 \\
Gl357 &  -0.28 & 3395 \\
Gl382 &  0.01 & 3672 \\
Gl393 &  -0.14 & 3569 \\
Gl402 &  0.02 & 3077 \\
Gl436 &  -0.04 & 3401 \\
Gl465 &  -0.56 & 3571 \\
Gl480.1 &  -0.54 & 3373 \\
Gl486 &  -0.01 & 3329 \\
Gl514 &  -0.15 & 3671 \\
Gl526 &  -0.15 & 3664 \\
Gl536 &  -0.12 & 3674 \\
Gl555 &  0.13 & 3258 \\
Gl581 &  -0.20 & 3344 \\
Gl628 &  -0.06 & 3362 \\
Gl643 &  -0.32 & 3243 \\
Gl678.1A &  -0.05 & 3704 \\
Gl686 &  -0.30 & 3670 \\
Gl699 &  -0.61 & 3256 \\
Gl701 &  -0.22 & 3646 \\
Gl752A &  0.02 & 3538 \\
Gl846 &  -0.07 & 3810 \\
Gl849 &  0.19 & 3516 \\
Gl876 &  0.13 & 3202 \\
Gl880 &  0.03 & 3703 \\
Gl908 &  -0.40 & 3475 \\
Gl1125 &  -0.15 & 3328 \\
Gl1129 &  0.00 & 3112 \\
Gl1256 &  -0.05 & 2992 \\
Gl1265 &  -0.21 & 3045 \\
Gl2066 &  -0.12 & 3597 \\
LP816-60 &  -0.11 & 3071 \\
Gl1 &  -0.42 & 3616 \\
Gl674 &  -0.19 & 3409 \\
Gl832 &  -0.21 & 3512 \\
Gl887 &  -0.29 & 3692 
\label{upgraderefparam}
\end{longtable}
\clearpage
%\end{appendix}
\twocolumn

\section{The determinations for 82 stars in SWEET-Cat}
\label{B}

Table~\ref{Bdet} contains the derived parameters with their maximum uncertainties, applying both the new and the old reference datasets. For stars with spectra from multiple instruments, we present all our measurements.

\onecolumn
\begin{longtable}{cccccc}

\caption[Determination for parameters and total error budgets for 82 stars in SWEET-Cat with the new and the old reference datasets]{Determinations for the parameters and total error budgets for 82 stars in SWEET-Cat with the new and the old reference datasets.} \\

\hline\hline \\
Star & Spec. & $T_{\mathrm{eff}}$ (New) & [Fe/H] (New) & $T_{\mathrm{eff}}$ (Old) & [Fe/H] (Old) \\
 & & [K] & [dex] & [K] & [dex]  \\
\hline \\
\endfirsthead
\caption{continued}\\
\hline\hline \\
Star & Spec. & $T_{\mathrm{eff}}$ (New) & [Fe/H] (New) & $T_{\mathrm{eff}}$ (Old) & [Fe/H] (Old) \\
 & & [K] & [dex] & [K] & [dex]  \\
\hline \\
\endhead

\hline \\
\endfoot
CD Cet & CARMENES & 3091$\pm$115 & -0.11$\pm$0.13 & 3113$\pm$98 & -0.23$\pm$0.14 \\
G264-012 & CARMENES & 3339$\pm$104 & -0.06$\pm$0.13 & 3237$\pm$87 & -0.12$\pm$0.13 \\
Gl15A & HARPS-N & 3602$\pm$91 & -0.45$\pm$0.11 & 3546$\pm$65 & -0.39$\pm$0.10 \\
Gl15A & CARMENES & 3656$\pm$107 & -0.33$\pm$0.13 & 3604$\pm$83 & -0.29$\pm$0.12 \\
Gl15A & SOPHIE & 3602$\pm$100 & -0.33$\pm$0.13 & 3484$\pm$78 & -0.30$\pm$0.13 \\
Gl27.1 & HARPS & 3715$\pm$90 & -0.14$\pm$0.11 & 3566$\pm$66 & -0.09$\pm$0.10 \\
Gl27.1 & UVES & 3713$\pm$94 & -0.11$\pm$0.11 & 3597$\pm$69 & -0.08$\pm$0.10 \\
Gl27.1 & FEROS & 3703$\pm$107 & -0.07$\pm$0.13 & 3622$\pm$87 & -0.06$\pm$0.13 \\ %
Gl49 & HARPS-N & 3726$\pm$91 & 0.00$\pm$0.11 & 3485$\pm$65 & 0.04$\pm$0.10 \\
Gl49 & CARMENES & 3670$\pm$100 & 0.03$\pm$0.12 & 3522$\pm$79 & 0.02$\pm$0.12 \\ %
Gl49 & SOPHIE & 3788$\pm$102 & 0.02$\pm$0.13 & 3438$\pm$80 & 0.12$\pm$0.13 \\
Gl96 & CARMENES & 3821$\pm$100 & -0.05$\pm$0.12 & 3680$\pm$80 & -0.04$\pm$0.13 \\
Gl96 & SOPHIE & 3830$\pm$100 & 0.00$\pm$0.13 & 3656$\pm$80 & -0.01$\pm$0.13 \\
Gl163 & HARPS & 3431$\pm$93 & 0.01$\pm$0.11 & 3217$\pm$65 & 0.01$\pm$0.10 \\
Gl163 & FEROS & 3507$\pm$108 & 0.03$\pm$0.13 & 3275$\pm$85 & 0.02$\pm$0.13 \\ %
Gl176 & HARPS & 3623$\pm$90 & -0.02$\pm$0.11 & 3390$\pm$65 & 0.01$\pm$0.10 \\
Gl176 & CARMENES & 3426$\pm$109 & -0.04$\pm$0.13 & 3389$\pm$90 & -0.13$\pm$0.13 \\
Gl176 & SOPHIE & 3628$\pm$102 & -0.03$\pm$0.13 & 3382$\pm$80 & 0.05$\pm$0.13 \\
Gl179 & CARMENES & 3420$\pm$103 & 0.13$\pm$0.12 & 3263$\pm$81 & 0.02$\pm$0.13 \\
Gl179 & SOPHIE & 3405$\pm$108 & 0.10$\pm$0.14 & 3140$\pm$81 & 0.08$\pm$0.13 \\
Gl179 & FEROS & 3386$\pm$116 & 0.06$\pm$0.13 & 3188$\pm$90 & 0.07$\pm$0.13 \\ %
Gl180 & HARPS & 3544$\pm$90 & -0.31$\pm$0.11 & 3423$\pm$65 & -0.26$\pm$0.10 \\
Gl180 & UVES & 3580$\pm$95 & -0.28$\pm$0.11 & 3485$\pm$71 & -0.28$\pm$0.10 \\
Gl180 & CARMENES & 3567$\pm$101 & -0.31$\pm$0.12 & 3501$\pm$80 & -0.27$\pm$0.12 \\
Gl180 & FEROS & 3585$\pm$105 & -0.24$\pm$0.13 & 3474$\pm$85 & -0.21$\pm$0.13 \\ %
Gl229 & HARPS & 3782$\pm$90 & -0.06$\pm$0.11 & 3577$\pm$65 & -0.01$\pm$0.10 \\
Gl229 & ESPRESSO & 3751$\pm$92 & -0.05$\pm$0.11 & 3555$\pm$66 & -0.02$\pm$0.10 \\
Gl229 & UVES & 3801$\pm$97 & -0.02$\pm$0.11 & 3622$\pm$71 & -0.04$\pm$0.10 \\
Gl229 & CARMENES & 3738$\pm$101 & -0.07$\pm$0.13 & 3605$\pm$79 & -0.05$\pm$0.12 \\
Gl251 & SOPHIE & 3397$\pm$100 & -0.05$\pm$0.13 & 3221$\pm$79 & -0.07$\pm$0.13 \\
Gl273 & HARPS & 3282$\pm$92 & -0.12$\pm$0.11 & 3103$\pm$65 & -0.09$\pm$0.10 \\
Gl273 & HARPS-N & 3297$\pm$93 & -0.15$\pm$0.11 & 3120$\pm$66 & -0.13$\pm$0.10 \\
Gl273 & CARMENES & 3237$\pm$107 & -0.19$\pm$0.13 & 3188$\pm$83 & -0.14$\pm$0.12 \\
Gl273 & SOPHIE & 3244$\pm$103 & -0.03$\pm$0.13 & 3153$\pm$80 & -0.08$\pm$0.13 \\ %
Gl273 & FEROS & 3344$\pm$137 & -0.20$\pm$0.14 & 3219$\pm$96 & -0.17$\pm$0.14 \\
Gl317 & HARPS & 3354$\pm$94 & 0.12$\pm$0.11 & 3143$\pm$66 & 0.14$\pm$0.10 \\
Gl317 & FEROS & 3497$\pm$122 & 0.12$\pm$0.14 & 3232$\pm$92 & 0.12$\pm$0.14 \\ %
Gl328 & UVES & 3925$\pm$98 & 0.06$\pm$0.11 & 3854$\pm$73 & -0.01$\pm$0.11 \\
Gl328 & SOPHIE & 3929$\pm$106 & 0.10$\pm$0.14 & 3829$\pm$84 & -0.01$\pm$0.14 \\ %
Gl328 & FEROS & 3924$\pm$112 & 0.15$\pm$0.14 & 3876$\pm$87 & -0.01$\pm$0.14 \\
Gl357 & HARPS & 3458$\pm$90 & -0.35$\pm$0.11 & 3342$\pm$65 & -0.30$\pm$0.10 \\
Gl357 & HARPS-N & 3473$\pm$94 & -0.37$\pm$0.11 & 3388$\pm$67 & -0.34$\pm$0.10 \\
Gl357 & UVES & 3474$\pm$94 & -0.29$\pm$0.11 & 3400$\pm$70 & -0.27$\pm$0.10 \\
Gl357 & FEROS & 3532$\pm$129 & -0.26$\pm$0.14 & 3426$\pm$100 & -0.24$\pm$0.14 \\
Gl367 & FEROS & 3651$\pm$108 & -0.03$\pm$0.13 & 3424$\pm$86 & -0.03$\pm$0.13 \\
Gl378 & SOPHIE & 3758$\pm$101 & 0.02$\pm$0.13 & 3613$\pm$80 & -0.01$\pm$0.13 \\
Gl393 & HARPS & 3549$\pm$92 & -0.20$\pm$0.11 & 3411$\pm$65 & -0.17$\pm$0.10 \\
Gl393 & ESPRESSO & 3530$\pm$92 & -0.22$\pm$0.11 & 3425$\pm$65 & -0.19$\pm$0.10 \\
Gl393 & CARMENES & 3561$\pm$102 & -0.18$\pm$0.12 & 3501$\pm$80 & -0.19$\pm$0.12 \\
Gl393 & SOPHIE & 3500$\pm$101 & -0.13$\pm$0.13 & 3419$\pm$80 & -0.15$\pm$0.13 \\
Gl393 & FEROS & 3628$\pm$106 & -0.15$\pm$0.13 & 3462$\pm$87 & -0.16$\pm$0.13 \\
Gl422 & HARPS & 3459$\pm$95 & -0.17$\pm$0.11 & 3259$\pm$66 & -0.16$\pm$0.10 \\
Gl422 & UVES & 3458$\pm$102 & -0.18$\pm$0.11 & 3308$\pm$73 & -0.20$\pm$0.11 \\
Gl422 & FEROS & 3527$\pm$118 & -0.15$\pm$0.13 & 3317$\pm$88 & -0.15$\pm$0.14 \\ %
Gl433 & HARPS & 3630$\pm$90 & -0.18$\pm$0.11 & 3468$\pm$65 & -0.14$\pm$0.10 \\
Gl433 & UVES & 3637$\pm$94 & -0.19$\pm$0.11 & 3521$\pm$67 & -0.19$\pm$0.10 \\
Gl436 & HARPS & 3478$\pm$91 & -0.08$\pm$0.11 & 3289$\pm$65 & -0.04$\pm$0.10 \\
Gl436 & HARPS-N & 3458$\pm$96 & -0.13$\pm$0.11 & 3326$\pm$67 & -0.12$\pm$0.10 \\
Gl436 & CARMENES & 3476$\pm$103 & -0.03$\pm$0.12 & 3418$\pm$84 & -0.09$\pm$0.12 \\
Gl436 & SOPHIE & 3422$\pm$100 & -0.01$\pm$0.13 & 3318$\pm$80 & -0.04$\pm$0.13 \\ %
Gl436 & FEROS & 3531$\pm$107 & -0.01$\pm$0.13 & 3347$\pm$84 & -0.04$\pm$0.13 \\
Gl463 & HARPS-N & 3532$\pm$95 & 0.05$\pm$0.11 & 3324$\pm$66 & 0.05$\pm$0.10 \\
Gl463 & CARMENES & 3582$\pm$109 & 0.02$\pm$0.13 & 3505$\pm$91 & -0.05$\pm$0.13 \\
Gl463 & SOPHIE & 3577$\pm$132 & 0.04$\pm$0.14 & 3339$\pm$96 & 0.06$\pm$0.14 \\
Gl486 & HARPS & 3246$\pm$93 & -0.06$\pm$0.11 & 3105$\pm$68 & -0.06$\pm$0.10 \\
Gl486 & ESPRESSO & 3231$\pm$112 & 0.06$\pm$0.13 & 3151$\pm$77 & 0.01$\pm$0.13 \\
Gl486 & UVES & 3340$\pm$97 & -0.06$\pm$0.11 & 3183$\pm$71 & -0.06$\pm$0.11 \\ %
Gl486 & FEROS & 3395$\pm$107 & -0.01$\pm$0.13 & 3173$\pm$85 & -0.06$\pm$0.13 \\
Gl536 & HARPS & 3694$\pm$90 & -0.16$\pm$0.11 & 3559$\pm$65 & -0.13$\pm$0.10 \\
Gl536 & HARPS-N & 3707$\pm$90 & -0.16$\pm$0.11 & 3583$\pm$65 & -0.16$\pm$0.10 \\
Gl536 & CARMENES & 3698$\pm$104 & -0.14$\pm$0.12 & 3634$\pm$80 & -0.17$\pm$0.12 \\ %
Gl536 & FEROS & 3706$\pm$124 & -0.10$\pm$0.14 & 3581$\pm$89 & -0.11$\pm$0.14 \\
Gl581 & HARPS & 3367$\pm$91 & -0.25$\pm$0.11 & 3224$\pm$65 & -0.21$\pm$0.10 \\
Gl581 & CARMENES & 3426$\pm$100 & -0.20$\pm$0.12 & 3359$\pm$85 & -0.19$\pm$0.12 \\ %
Gl581 & FEROS & 3468$\pm$115 & -0.17$\pm$0.13 & 3305$\pm$90 & -0.21$\pm$0.14 \\
Gl625 & HARPS-N & 3509$\pm$92 & -0.51$\pm$0.11 & 3459$\pm$66 & -0.47$\pm$0.10 \\
Gl625 & CARMENES & 3530$\pm$106 & -0.45$\pm$0.12 & 3487$\pm$80 & -0.42$\pm$0.12 \\
Gl625 & SOPHIE & 3548$\pm$102 & -0.41$\pm$0.13 & 3419$\pm$84 & -0.36$\pm$0.13 \\ %
Gl649 & HARPS & 3720$\pm$90 & -0.09$\pm$0.11 & 3546$\pm$65 & -0.04$\pm$0.10 \\
Gl649 & CARMENES & 3671$\pm$102 & -0.06$\pm$0.12 & 3577$\pm$69 & -0.06$\pm$0.12 \\
Gl649 & SOPHIE & 3708$\pm$100 & -0.05$\pm$0.13 & 3512$\pm$80 & -0.04$\pm$0.13 \\ %
Gl674 & HARPS & 3425$\pm$91 & -0.26$\pm$0.11 & 3304$\pm$65 & -0.22$\pm$0.10 \\
Gl674 & ESPRESSO & 3417$\pm$92 & -0.29$\pm$0.11 & 3292$\pm$66 & -0.24$\pm$0.10 \\
Gl674 & FEROS & 3484$\pm$108 & -0.16$\pm$0.13 & 3344$\pm$84 & -0.17$\pm$0.13 \\
Gl676A & HARPS & 3978$\pm$92 & 0.08$\pm$0.11 & 3732$\pm$67 & 0.08$\pm$0.11 \\
Gl676A & FEROS & 3982$\pm$114 & 0.12$\pm$0.14 & 3775$\pm$88 & 0.08$\pm$0.14 \\ %
Gl680 & HARPS & 3572$\pm$90 & -0.22$\pm$0.11 & 3431$\pm$65 & -0.17$\pm$0.10 \\
Gl680 & FEROS & 3603$\pm$109 & -0.24$\pm$0.13 & 3473$\pm$87 & -0.13$\pm$0.14 \\
Gl682 & UVES & 3202$\pm$101 & -0.02$\pm$0.11 & 3108$\pm$73 & -0.06$\pm$0.10 \\
Gl682 & ESPRESSO & 3232$\pm$106 & 0.08$\pm$0.12 & 3102$\pm$74 & 0.05$\pm$0.11 \\
Gl682 & FEROS & 3324$\pm$112 & 0.01$\pm$0.14 & 3100$\pm$92 & 0.01$\pm$0.14 \\ %
Gl685 & HARPS-N & 3787$\pm$91 & -0.08$\pm$0.11 & 3635$\pm$65 & -0.04$\pm$0.10 \\
Gl685 & CARMENES & 3721$\pm$103 & -0.05$\pm$0.12 & 3642$\pm$80 & -0.04$\pm$0.12 \\
Gl685 & SOPHIE & 3754$\pm$101 & -0.05$\pm$0.13 & 3614$\pm$81 & -0.01$\pm$0.13 \\ %
Gl686 & HARPS & 3632$\pm$91 & -0.35$\pm$0.11 & 3566$\pm$65 & -0.31$\pm$0.10 \\
Gl686 & CARMENES & 3633$\pm$103 & -0.26$\pm$0.13 & 3563$\pm$82 & -0.29$\pm$0.13 \\ %
Gl686 & SOPHIE & 3657$\pm$101 & -0.28$\pm$0.13 & 3577$\pm$79 & -0.27$\pm$0.13 \\
Gl686 & FEROS & 3652$\pm$108 & -0.22$\pm$0.13 & 3612$\pm$88 & -0.24$\pm$0.14 \\ %
Gl687 & HARPS-N & 3453$\pm$94 & -0.03$\pm$0.11 & 3273$\pm$66 & -0.03$\pm$0.10 \\
Gl687 & SOPHIE & 3453$\pm$104 & 0.03$\pm$0.13 & 3288$\pm$82 & 0.00$\pm$0.13 \\ %
Gl699 & HARPS & 3226$\pm$91 & -0.62$\pm$0.11 & 3113$\pm$68 & -0.58$\pm$0.10 \\
Gl699 & HARPS-N & 3258$\pm$93 & -0.62$\pm$0.11 & 3160$\pm$68 & -0.59$\pm$0.10 \\
Gl699 & UVES & 3300$\pm$97 & -0.57$\pm$0.11 & 3191$\pm$75 & -0.56$\pm$0.11 \\
Gl699 & CARMENES & 3228$\pm$114 & -0.53$\pm$0.13 & 3161$\pm$87 & -0.58$\pm$0.13 \\
Gl699 & SOPHIE & 3199$\pm$103 & -0.52$\pm$0.13 & 3161$\pm$82 & -0.53$\pm$0.13 \\
Gl720A & HARPS-N & 3832$\pm$92 & -0.07$\pm$0.11 & 3759$\pm$66 & -0.12$\pm$0.10 \\
Gl720A & CARMENES & 3778$\pm$106 & -0.06$\pm$0.13 & 3773$\pm$81 & -0.12$\pm$0.13 \\ %
Gl720A & SOPHIE & 3774$\pm$103 & 0.05$\pm$0.13 & 3716$\pm$81 & -0.04$\pm$0.13 \\
Gl740 & HARPS & 3832$\pm$91 & -0.04$\pm$0.11 & 3663$\pm$66 & -0.02$\pm$0.10 \\
Gl740 & HARPS-N & 3837$\pm$91 & -0.06$\pm$0.11 & 3691$\pm$66 & -0.04$\pm$0.10 \\
Gl740 & CARMENES & 3779$\pm$103 & -0.01$\pm$0.13 & 3721$\pm$82 & -0.05$\pm$0.13 \\
Gl740 & SOPHIE & 3824$\pm$102 & -0.02$\pm$0.13 & 3717$\pm$84 & -0.05$\pm$0.13 \\
Gl740 & FEROS & 3808$\pm$102 & 0.02$\pm$0.13 & 3673$\pm$82 & -0.01$\pm$0.13 \\
Gl752A & HARPS & 3554$\pm$91 & 0.02$\pm$0.11 & 3337$\pm$65 & 0.05$\pm$0.10 \\
Gl752A & CARMENES & 3540$\pm$100 & 0.01$\pm$0.12 & 3384$\pm$78 & 0.02$\pm$0.12 \\
Gl752A & SOPHIE & 3580$\pm$99 & 0.09$\pm$0.13 & 3349$\pm$80 & 0.05$\pm$0.13 \\ %
Gl832 & HARPS & 3587$\pm$90 & -0.18$\pm$0.11 & 3459$\pm$65 & -0.14$\pm$0.10 \\
Gl832 & FEROS & 3563$\pm$103 & -0.11$\pm$0.13 & 3466$\pm$82 & -0.11$\pm$0.13 \\
Gl849 & HARPS & 3486$\pm$95 & 0.16$\pm$0.11 & 3208$\pm$66 & 0.20$\pm$0.10 \\
Gl849 & CARMENES & 3471$\pm$106 & 0.20$\pm$0.12 & 3333$\pm$89 & 0.12$\pm$0.12 \\
Gl876 & FEROS & 3265$\pm$108 & 0.07$\pm$0.13 & 3102$\pm$84 & 0.08$\pm$0.13 \\
Gl887 & HARPS & 3679$\pm$90 & -0.23$\pm$0.11 & 3578$\pm$65 & -0.19$\pm$0.10 \\
Gl887 & FEROS & 3701$\pm$112 & -0.17$\pm$0.14 & 3584$\pm$84 & -0.14$\pm$0.13 \\
Gl1214 & UVES & 3109$\pm$129 & -0.12$\pm$0.12 & 3005$\pm$80 & -0.06$\pm$0.12 \\
Gl1214 & FEROS & 3115$\pm$129 & -0.04$\pm$0.14 & 3022$\pm$95 & -0.09$\pm$0.14 \\ %
Gl3082 & HARPS & 3645$\pm$90 & -0.23$\pm$0.11 & 3508$\pm$65 & -0.18$\pm$0.10 \\
Gl3082 & UVES & 3674$\pm$95 & -0.19$\pm$0.11 & 3570$\pm$69 & -0.17$\pm$0.10 \\
Gl3082 & FEROS & 3681$\pm$108 & -0.14$\pm$0.13 & 3550$\pm$83 & -0.12$\pm$0.13 \\ %
Gl3090 & HARPS & 3701$\pm$91 & -0.02$\pm$0.11 & 3509$\pm$67 & 0.01$\pm$0.10 \\
Gl3090 & FEROS & 3737$\pm$104 & 0.04$\pm$0.13 & 3536$\pm$83 & 0.02$\pm$0.13 \\ %
Gl3138 & HARPS & 3884$\pm$92 & 0.02$\pm$0.11 & 3797$\pm$69 & -0.05$\pm$0.11 \\
Gl3323 & FEROS & 3195$\pm$114 & -0.34$\pm$0.14 & 3092$\pm$89 & -0.33$\pm$0.14 \\
Gl3341 & HARPS & 3455$\pm$100 & 0.08$\pm$0.12 & 3433$\pm$74 & -0.02$\pm$0.12 \\
Gl3470 & UVES & 3620$\pm$106 & 0.03$\pm$0.11 & 3385$\pm$76 & 0.09$\pm$0.11 \\
Gl3634 & HARPS & 3535$\pm$92 & -0.08$\pm$0.11 & 3347$\pm$65 & -0.02$\pm$0.10 \\
Gl3942 & HARPS-N & 3823$\pm$90 & -0.04$\pm$0.11 & 3680$\pm$66 & -0.04$\pm$0.10 \\
Gl3998 & HARPS & 3729$\pm$91 & -0.13$\pm$0.11 & 3557$\pm$65 & -0.09$\pm$0.10 \\
Gl3998 & HARPS-N & 3720$\pm$91 & -0.13$\pm$0.11 & 3557$\pm$65 & -0.07$\pm$0.10 \\
Gl3998 & FEROS & 3712$\pm$102 & -0.03$\pm$0.13 & 3546$\pm$83 & -0.03$\pm$0.13 \\ %
Gl9066 & CARMENES & 3317$\pm$114 & -0.20$\pm$0.13 & 3232$\pm$102 & -0.35$\pm$0.14 \\
Gl9689 & HARPS & 3832$\pm$94 & 0.08$\pm$0.11 & 3674$\pm$66 & 0.05$\pm$0.10 \\
Gl9689 & HARPS-N & 3850$\pm$94 & 0.01$\pm$0.11 & 3692$\pm$66 & 0.00$\pm$0.10 \\
HD238090 & CARMENES & 3812$\pm$107 & -0.03$\pm$0.13 & 3800$\pm$83 & -0.11$\pm$0.13 \\
HD260655 & ESPRESSO & 3639$\pm$103 & -0.21$\pm$0.12 & 3632$\pm$69 & -0.22$\pm$0.11 \\
HD260655 & CARMENES & 3671$\pm$116 & -0.21$\pm$0.14 & 3750$\pm$82 & -0.28$\pm$0.13 \\
HIP4845 & HARPS & 3915$\pm$98 & 0.10$\pm$0.12 & 3900$\pm$76 & -0.06$\pm$0.13 \\
HIP36985 & HARPS & 3809$\pm$95 & -0.01$\pm$0.11 & 3634$\pm$72 & -0.01$\pm$0.11 \\
HIP36985 & CARMENES & 3800$\pm$105 & 0.03$\pm$0.13 & 3667$\pm$87 & -0.05$\pm$0.13 \\ %
HIP36985 & FEROS & 3834$\pm$110 & 0.05$\pm$0.14 & 3643$\pm$93 & 0.01$\pm$0.14 \\ %
HIP38594 & HARPS & 3903$\pm$97 & 0.03$\pm$0.11 & 3874$\pm$72 & -0.10$\pm$0.12 \\
HIP57050 & CARMENES & 3305$\pm$102 & 0.06$\pm$0.12 & 3161$\pm$80 & 0.02$\pm$0.12 \\
HIP57050 & SOPHIE & 3348$\pm$104 & 0.03$\pm$0.13 & 3134$\pm$82 & 0.01$\pm$0.13 \\
HIP79431 & FEROS & 3488$\pm$124 & 0.10$\pm$0.13 & 3281$\pm$89 & 0.20$\pm$0.13 \\
K2-3 & HARPS & 3788$\pm$96 & -0.11$\pm$0.12 & 3815$\pm$69 & -0.19$\pm$0.11 \\
K2-3 & HARPS-N & 3752$\pm$96 & -0.09$\pm$0.12 & 3773$\pm$70 & -0.16$\pm$0.11 \\
K2-3 & FEROS & 3748$\pm$117 & -0.04$\pm$0.14 & 3836$\pm$94 & -0.16$\pm$0.14 \\ %
K2-18 & CARMENES & 3587$\pm$113 & -0.09$\pm$0.13 & 3534$\pm$112 & -0.19$\pm$0.15 \\
K2-122 & HARPS-N & 3906$\pm$94 & 0.05$\pm$0.11 & 3668$\pm$66 & 0.03$\pm$0.10 \\
L98-59 & HARPS & 3420$\pm$92 & -0.34$\pm$0.11 & 3297$\pm$66 & -0.31$\pm$0.10 \\
L168-9 & HARPS & 3872$\pm$91 & 0.05$\pm$0.11 & 3634$\pm$65 & 0.06$\pm$0.10 \\
L168-9 & FEROS & 3842$\pm$103 & 0.11$\pm$0.13 & 3645$\pm$84 & 0.10$\pm$0.13 \\ %
Lalande21185 & UVES & 3529$\pm$115 & -0.26$\pm$0.12 & 3467$\pm$76 & -0.23$\pm$0.11 \\
Lalande21185 & CARMENES & 3518$\pm$121 & -0.32$\pm$0.12 & 3557$\pm$81 & -0.35$\pm$0.12 \\ %
LHS1678 & HARPS & 3549$\pm$94 & -0.45$\pm$0.11 & 3456$\pm$65 & -0.42$\pm$0.10 \\
LHS1815 & HARPS & 3678$\pm$92 & 0.05$\pm$0.11 & 3458$\pm$66 &  0.05$\pm$0.10 \\
LP714-47 & ESPRESSO & 3860$\pm$92 & 0.06$\pm$0.11 & 3606$\pm$66 & 0.08$\pm$0.10 \\
LSPMJ2116+0234 & FEROS & 3590$\pm$108 & -0.03$\pm$0.13 & 3302$\pm$86 & -0.01$\pm$0.13 \\
LTT1445A & HARPS & 3374$\pm$92 & -0.36$\pm$0.11 & 3240$\pm$68 & -0.35$\pm$0.10 \\
LTT1445A & ESPRESSO & 3330$\pm$92 & -0.42$\pm$0.11 & 3242$\pm$68 & -0.39$\pm$0.10 \\
Proxima Centauri & HARPS & 2756$\pm$112 & -0.18$\pm$0.11 & 2702$\pm$69 & -0.12$\pm$0.10 \\
Proxima Centauri & UVES & 2804$\pm$113 & -0.21$\pm$0.11 & 2833$\pm$78 & -0.23$\pm$0.11 \\ %
Ross128 & HARPS & 3114$\pm$93 & -0.30$\pm$0.11 & 2983$\pm$66 & -0.26$\pm$0.10 \\
Ross128 & FEROS & 3166$\pm$111 & -0.30$\pm$0.13 & 3085$\pm$90 & -0.28$\pm$0.14 \\
TOI-270 & HARPS & 3539$\pm$92 & -0.31$\pm$0.11 & 3422$\pm$66 & -0.29$\pm$0.10 \\
TOI-270 & ESPRESSO & 3465$\pm$93 & -0.38$\pm$0.11 & 3407$\pm$66 & -0.33$\pm$0.10 \\
TOI-776 & HARPS & 3765$\pm$90 & -0.11$\pm$0.11 & 3634$\pm$66 & -0.10$\pm$0.10 \\
TOI-776 & ESPRESSO & 3730$\pm$91 & -0.13$\pm$0.11 & 3621$\pm$66 & -0.12$\pm$0.10 \\
TOI-1235 & HARPS-N & 3925$\pm$94 & 0.03$\pm$0.11 & 3824$\pm$68 & -0.04$\pm$0.11 \\
TOI-1266 & HARPS-N & 3614$\pm$92 & -0.32$\pm$0.11 & 3521$\pm$66 & -0.28$\pm$0.10 \\
TOI-1801 & HARPS-N & 3797$\pm$91 & -0.07$\pm$0.11 & 3698$\pm$69 & -0.09$\pm$0.11 \\
TYC-2187-512-1 & CARMENES & 3682$\pm$102 & -0.16$\pm$0.13 & 3620$\pm$80 & -0.14$\pm$0.13 \\
TYC-2187-512-1 & FEROS & 3678$\pm$105 & -0.16$\pm$0.13 & 3615$\pm$84 & -0.14$\pm$0.13 \\ %
Wolf1061 & HARPS & 3307$\pm$92 & -0.12$\pm$0.11 & 3103$\pm$66 & -0.08$\pm$0.10 \\
Wolf1061 & CARMENES & 3315$\pm$100 & -0.03$\pm$0.12 & 3211$\pm$81 & -0.07$\pm$0.12 \\ %
Wolf1061 & FEROS & 3357$\pm$132 & -0.10$\pm$0.14 & 3207$\pm$99 & -0.10$\pm$0.14  %
\label{Bdet}
\end{longtable}
%\end{appendix}

\end{appendix}

\end{document}